\documentclass[aps,pra,twocolumn,showpacs]{revtex4-1}
\usepackage{epsfig}
\usepackage{bm}

\newcommand{\be}{\begin{eqnarray}}
\newcommand{\ee}{\end{eqnarray}}
\newcommand{\ba}{\begin{array}}
\newcommand{\ea}{\end{array}}
\newcommand{\bmat}{\left(\begin{array}}
\newcommand{\emat}{\end{array}\right)}
\newcommand{\no}{\nonumber}

\newcommand{\e}{\mathrm e}

\begin{document}
\title{Shortcuts to adiabaticity for quantum annealing}
\author{Kazutaka Takahashi}
\affiliation{Department of Physics, Tokyo Institute of Technology, 
Tokyo 152-8551, Japan}

\date{\today}

\begin{abstract} 
We study the Ising Hamiltonian with a transverse field term to 
simulate the quantum annealing.
Using shortcuts to adiabaticity, we design 
the time dependence of the Hamiltonian.
The dynamical invariant is obtained by the mean-field ansatz, 
and the Hamiltonian is designed by the inverse engineering.
We show that the time dependence of physical quantities 
such as the magnetization is independent of the speed of 
the Hamiltonian variation in the infinite-range model.
We also show that rotating transverse magnetic fields are 
useful to achieve the ideal time evolution.
\end{abstract}
\maketitle

\section{Introduction}

Quantum annealing is a quantum algorithm that solves 
combinatorial optimization problems~\cite{KN,BBRA, FGGS, FGGLLP}.
The solution of the problem is described by the ground state
of an Ising-spin Hamiltonian, 
and the transverse field is introduced to find the state 
by the quantum time evolution.
The statistical properties of the transverse Ising model 
have been discussed for decades in many contexts.
The appearance of hardware realizing quantum annealing 
has attracted renewed interest~\cite{BRIWWLMT}  
and accelerated active studies among researchers in various fields.

The principle of the algorithm is based on 
the quantum mechanical time evolution.
Theoretically, the method requires adiabatic time evolutions and
the system Hamiltonian must be changed infinitely slowly.
In realistic situations with a finite speed, 
the system undergoes nonadiabatic transitions, and it is required that 
their losses are minimized to obtain effective solutions.

A possible improvement is to optimize the time dependence of the Hamiltonian.
According to the adiabatic approximation, 
the error is reduced by considering a slower change 
of the Hamiltonian at initial and final times~\cite{Morita, LRH, WB}.
It was demonstrated that this boundary cancellation method improves 
the performance of computing even in open systems~\cite{AL}. 
Another possible improvement is to change the driver Hamiltonian 
from the transverse field term to other different operators.
This possibility has been discussed intensively in recent studies.
It is discussed that the ``nonstoquastic'' Hamiltonian improves
the performance of quantum annealing~\cite{BDOT, FGG, SN, CFLLS, HBCT}.

The problem for their improvements is that 
there are no general principles to determine the best algorithm 
systematically.
In this paper, we consider the Hamiltonian design
using the idea of shortcuts to adiabaticity~\cite{DR1, DR2, Berry, CRSCGM, STA}.
The Hamiltonian is constructed so that    
the system follows an adiabatic passage of a reference Hamiltonian 
with arbitrary speed. 

There are several ways to implement the method to given systems.
In the counterdiabatic driving, we introduce the counterdiabatic term
to prevent the nonadiabatic transitions.
Then the system follows instantaneous eigenstates of the original Hamiltonian.
This method requires an additional term to the Hamiltonian.
The counterdiabatic term often takes a complicated form 
and is generally difficult to be realized in experiments.
This problem does not occur in the method of inverse engineering.
The operator form of the Hamiltonian is kept fixed and 
the time dependence of the coefficients is determined 
so that we can find the desired time evolutions.
This is suitable to the present problem of quantum annealing
and we use this method in the present study.
By solving the equation for the dynamical invariant~\cite{LR},
we can design the Hamiltonian.
We can also find a similar result by considering a unitary transformation
of the Hamiltonian of the counterdiabatic driving~\cite{Takahashi1}.

The design of the Hamiltonian for one-dimensional 
many-spin systems has been discussed in many works.
The one-dimensional Ising and $XY$ spin chains are exactly solvable 
and the counterdiabatic term is constructed exactly and 
approximately~\cite{dCZ, Takahashi2, SOMdC, Damski, SP, OT}.
They are only applied to one-dimensional systems and 
the generalization is considered to be a formidable task.
In the opposite limit of the infinite-range systems,
the mean-field ansatz gives the exact result. 
The infinite-range $XY$ model called 
the Lipkin-Meshkov-Glick model 
was analyzed in Refs.~\cite{Takahashi2, CCPPF}, 
but the Ising system has not been considered before.
In addition to that, the inverse engineering has not been considered 
for many-spin systems.

In this paper, we consider the inverse engineering for 
the transverse Ising-spin model.
We use the mean-field ansatz to solve the problem.
We also discuss modifying the driver Hamiltonian.
For a reason that we discuss below, we treat a rotating transverse 
magnetic field.
We show that the rotating field 
is very convenient for quantum annealing.
The Hamiltonian with a rotating magnetic field is also interpreted 
as a nonstoquastic system and 
the nonstoquastic effect can also be studied in the present work.

The organization of the paper is as follows.
In Sec.~\ref{sec:di}, we briefly review the principle 
of shortcuts to adiabaticity.
The dynamical invariant plays an important role there, and
we give the basic properties.
The method is applied to single-spin systems in Sec.~\ref{sec:single}.
The solution is directly applied to 
many-spin systems in Sec.~\ref{sec:ising}. 
We discuss possible applications to transverse Ising model.
Section~\ref{sec:conc} is devoted to conclusions.

\section{Dynamical invariant}
\label{sec:di}

The method of shortcuts to adiabaticity 
is based on the equation for the dynamical invariant~\cite{CRSCGM,LR}.
For a time-dependent Hamiltonian $\hat{H}(t)$, 
the dynamical invariant operator $\hat{F}(t)$ satisfies 
the equation of motion: 
\be
 i\frac{\partial\hat{F}(t)}{\partial t}
 = [\hat{H}(t),\hat{F}(t)]. \label{LR}
\ee
This has the same form as the von Neumann equation.
$\hat{F}(t)$ includes the density operator as an example, 
but it does not need to be a positive operator in general.

When the Hermitian operator $\hat{F}(t)$ satisfies Eq.~(\ref{LR}), 
we can show the following properties.
First, eigenvalues of $\hat{F}(t)$ are time independent: 
\be
 \hat{F}(t) = \sum_n g_n|n(t)\rangle\langle n(t)|.
\ee
The eigenvalue $g_n$ can take arbitrary real values 
if we do not impose any additional constraints.
$|n(t)\rangle$ denotes the time-dependent eigenstate and 
is used to show the following second property.
The state, satisfying the Schr\"odinger equation 
for the Hamiltonian $\hat{H}(t)$, is written as 
\be
 |\psi(t)\rangle = \sum_n c_n\e^{i\alpha_n(t)}|n(t)\rangle.
\ee
The point is that the coefficient of $|n(t)\rangle$ except the phase, 
$c_n$, is time independent. 
This means that the state remains the same eigenstate throughout 
the time evolution.
Third, the Hamiltonian takes the form  
\be
 \hat{H}(t)&=& \sum_n E_n(t)|n(t)\rangle\langle n(t)| \no\\
 && +i\sum_n \left(1-|n(t)\rangle\langle n(t)|\right)
 |\dot{n}(t)\rangle\langle n(t)|,
\ee
where $E_n(t)$ represents a real function 
and the dot symbol denotes the time derivative.
The first term defines the adiabatic state.
The solution of the Schr\"odinger equation is given by 
the adiabatic state defined from the first term.
The second term plays the role of preventing nonadiabatic transitions.
For a given first term, we can realize the adiabatic time evolution in
a finite duration by introducing the second term.
The second term is called the counterdiabatic term.

This method of counterdiabatic driving works well 
if we can find the form of the counterdiabatic term.
It is usually difficult to find the explicit form 
of the counterdiabatic term in many-body systems 
because we need to know all eigenstates of the adiabatic Hamiltonian.
In addition, the counterdiabatic term takes a complicated form and 
is difficult to be realized in experiments.
To avoid these problems, 
we use the method of inverse engineering in the following.

\section{Inverse engineering for single spin system}
\label{sec:single}

We describe the method of inverse engineering 
by using a single-spin Hamiltonian
\be
 \hat{H}(t)= \bm{h}(t)\cdot\hat{\bm{S}},  \label{single}
\ee
where $\hat{\bm{S}}$ denotes spin-$\frac{1}{2}$ operators
and $\bm{h}(t)$ is a time-dependent magnetic field.
Although this system was studied in previous works,
we closely look at this simple model 
to apply the result directly to many-body systems
in the next section.

As an example, we take the field in $zx$ plane as
\be
 \bm{h}(t)=(\Gamma(t), 0, h_z(t)).
\ee
The state satisfies the Schr\"odinger equation 
\be
 i\frac{\partial}{\partial t}|\psi(t)\rangle
 = \hat{H}(t)|\psi(t)\rangle.
\ee
For a specified initial state $|\psi(0)\rangle=|\psi_{\rm I}\rangle$ 
and a final one $|\psi(T)\rangle=|\psi_{\rm F}\rangle$, 
we want to design the time dependence of the magnetic field $\bm{h}(t)$.

For the Hamiltonian (\ref{single}),
we can easily solve the equation for the dynamical invariant.
We set the invariant as 
\be
 \hat{F}(t) = \bm{n}(t)\cdot\hat{\bm{S}}, \label{Fsingle}
\ee
where $\bm{n}(t)$ represents a unit vector.
Since Eq.~(\ref{LR}) is invariant under 
the time-independent scale transformation $\hat{F}(t)\to\lambda\hat{F}(t)$,
we set $\bm{n}(t)$ as a unit vector.
The instantaneous eigenvalues of $\hat{F}(t)$ are given by time-independent 
constants $\pm 1/2$ as we require from the general property. 
Substituting Eq.~(\ref{Fsingle}) to (\ref{LR}), and using
the commutation relation
\be
 [\hat{S}^\mu,\hat{S}^\nu] = i\sum_\lambda
 \epsilon_{\mu\nu\lambda}\hat{S}^\lambda,
\ee
we obtain the equation of motion for the classical spin: 
\be
 \dot{\bm{n}}(t)=\bm{h}(t)\times\bm{n}(t). \label{eomsingle}
\ee

We usually solve the equation of motion (\ref{eomsingle})
for a given magnetic field function $\bm{h}(t)$.
In the inverse engineering, we design the magnetic field
by specifying a proper $\bm{n}(t)$.
We parametrize the unit vector
\be
 \bm{n}(t)=(
 \sin\theta(t)\cos\varphi(t), \sin\theta(t)\sin\varphi(t), \cos\theta(t)).
 \label{n}
\ee
Then, Eq.~(\ref{eomsingle}) is written as
\be
 && \Gamma(t)=\frac{\dot{n}_3(t)}{n_2(t)}=-\frac{\dot{\theta}(t)}{\sin\varphi(t)}, \label{eq1}\\
 && h_z(t)=-\frac{\dot{n}_1(t)}{n_2(t)}
 =-\dot{\theta}(t)\frac{\cos\theta(t)\cos\varphi(t)}{\sin\theta(t)\sin\varphi(t)}
 +\dot{\varphi}(t). 
 \label{eq2}
\ee
For a specified $(\theta(t),\varphi(t))$, 
the magnetic field is obtained from these equations.
We note that the state follows the adiabatic passage defined by the invariant.
This is not the eigenstate of the Hamiltonian.
Therefore, we impose the condition that
the state becomes the eigenstate of the Hamiltonian 
at initial and final times.
This means that $\hat{H}(t)$ and $\hat{F}(t)$ commute with each other
at $t=0$ and $t=T$:
\be
 [\hat{H}(0),\hat{F}(0)]=[\hat{H}(T),\hat{F}(T)]=0.
\ee

As an example, we consider the condition 
\be
 && (h_z(0),\Gamma(0))=(0,\Gamma_0), \label{b1}\\
 && (h_z(T),\Gamma(T))=(h_{1},0). \label{b2}
\ee
The magnetic field is in $x$ direction at $t=0$,
and in $z$ direction at $t=T$.
This means that we require the conditions
\be
 && (\theta(0),\varphi(0))=(\pi/2,0), \label{a-b1}\\
 && \theta(T)=0. \label{a-b2}
\ee
$\varphi(T)$ is left undetermined.
From the equations of motion, we also need additional conditions: 
\be
 && (\dot{\theta}(0),\dot{\varphi}(0))=(0,0), \label{a-b3}\\
 && (\dot{\theta}(T), \varphi(T))=(0,\pi/2). \label{a-b4}
\ee
The angle functions are not unique and we can choose
any functions which satisfy 
the boundary conditions~(\ref{a-b1})--(\ref{a-b4}).
For specified angles, we can determine the time dependence of 
the magnetic field from Eqs.~(\ref{eq1}) and (\ref{eq2}).
The details are described in Appendix~\ref{app:single}.

\section{Inverse engineering for transverse Ising model}
\label{sec:ising}

\subsection{Mean-field ansatz}

We apply the inverse engineering to many-spin systems.
To use the mean-field approximation, 
we consider the infinite-range Ising model in a transverse field, 
\be
 \hat{H}(t) &=&
 f(t)\left(
 -\frac{J}{2N}\sum_{i,j=1}^N\hat{\sigma}_i^z\hat{\sigma}_j^z
 -h\sum_{i=1}^N\hat{\sigma}_i^z\right) \no\\
 && -\Gamma(t)\sum_{i=1}^N\hat{\sigma}_i^x, \label{H}
\ee
where $\hat{\sigma}_i^z$ and $\hat{\sigma}_i^x$ are 
Pauli matrices at site $i$.
The number of spins is denoted by $N$ and is taken to be a large value.
The coupling $J$ denotes the interaction between two spins, 
and $h$ is the magnetic field in $z$ direction.
This is a typical Hamiltonian for quantum annealing.
Our purpose is to determine the time dependence of 
real functions $f(t)$ and $\Gamma(t)$
under the initial condition $(f(0),\Gamma(0))=(0,\Gamma_0)$
and the final condition $(f(T),\Gamma(T))=(1,0)$.

It is a difficult problem to solve the equation of motion (\ref{LR}).
Here we expect that the mean-field ansatz works 
well for the present problem.
For the same form of the Hamiltonian without time dependence,
it is well known that the mean-field ansatz gives the exact result 
as described in a standard textbook of statistical mechanics~\cite{NO}.
In the infinite-range model, a single spin interacts with the other spins
and the law of large numbers implies that the interaction is represented 
by an effective magnetic field.
We solve the equation by the mean-field ansatz
to examine whether the same property holds for dynamical systems.

In mean-field system, the Hamiltonian is replaced 
by a noninteracting one-body Hamiltonian.
Following the previous section, we put the form of 
the dynamical invariant as 
\be
 \hat{F}(t)=\bm{n}(t)\cdot\sum_{i=1}^N \hat{\bm{\sigma}}_i,
 \label{LRn}
\ee
where $\bm{n}(t)$ denotes a unit vector.
Then we impose Eq.~(\ref{LR}).
The commutator on the left-hand side is calculated to give
\be
 && [\hat{H}(t),\hat{F}(t)] \no\\
 &=& -\frac{if(t)J}{N}\sum_{i,j=1}^N\left[
 (\bm{n}(t)\times\hat{\bm{\sigma}}_i)^z\hat{\sigma}_j^z
 +\hat{\sigma}_i^z(\bm{n}(t)\times\hat{\bm{\sigma}}_j)^z
 \right] \no\\
 && -2i\sum_{i=1}^N \left[
 f(t)h(\bm{n}(t)\times\hat{\bm{\sigma}}_i)^z
 +\Gamma(t)(\bm{n}(t)\times\hat{\bm{\sigma}}_i)^x\right]. 
\ee
The first term on the right-hand side includes two Pauli operators, 
which prevents from satisfying Eq.~(\ref{LR}).
We use the mean-field ansatz
\be
 && (\bm{n}\times\hat{\bm{\sigma}}_i)^z\hat{\sigma}_j^z
 +\hat{\sigma}_i^z(\bm{n}\times\hat{\bm{\sigma}}_j)^z \no\\
 &\sim& 
 (\bm{n}\times\langle\hat{\bm{\sigma}}_i\rangle)^z\hat{\sigma}_j^z
 +\hat{\sigma}_i^z(\bm{n}\times\langle\hat{\bm{\sigma}}_j\rangle)^z
 +(\bm{n}\times\hat{\bm{\sigma}}_i)^z\langle\hat{\sigma}_j^z\rangle \no\\
 && +\langle\hat{\sigma}_i^z\rangle(\bm{n}
 \times\hat{\bm{\sigma}}_j)^z, \label{mfa}
\ee
where the bracket denotes the average in terms of 
the state $|\psi(t)\rangle$.
One of the Pauli operators is replaced with the average 
in terms of the state at each time.
In the present ansatz, the invariant is given by Eq.~(\ref{LRn}).
This means that the state $|\psi(t)\rangle$ is the coherent state
denoted by the vector $\bm{n}(t)$.
We then have 
\be
 \langle\hat{\bm{\sigma}}\rangle = \bm{n}(t), 
\ee
and the replacement (\ref{mfa}) gives 
the equation of motion for the unit vector $\bm{n}(t)$ as 
\be
 \dot{\bm{n}}(t) \sim -2f(t)(Jn_z(t)+h)\bm{e}_z\times\bm{n}(t)
 -2\Gamma(t) \bm{e}_x\times\bm{n}(t). \no\\
\ee
The parametrization in Eq.~(\ref{n}) gives 
\be
 && \Gamma(t) = \frac{\dot{\theta}(t)}{2\sin\varphi(t)}, \label{gamma}\\
 && f(t) = \frac{1}{2(J\cos\theta(t)+h)}\left(
 2\Gamma(t)\frac{\cos\theta(t)}{\sin\theta(t)}\cos\varphi(t)
 -\dot{\varphi}(t)\right). \no\\ \label{f}
\ee
These equations are very similar to Eqs.~(\ref{eq1}) and (\ref{eq2}) and 
are obtained by using the replacement:  
\be
 && \varphi \to -\varphi, \\
 && \Gamma \to 2\Gamma, \\
 && h \to 2f(J\cos\theta+h).
\ee

In quantum annealing, we usually set the initial and final conditions as 
\be
 && (f(0),\Gamma(0))=(0,\Gamma_0), \label{initial}\\
 && (f(T),\Gamma(T)=(1,0). \label{final}
\ee
We determine $\theta(t)$ and $\varphi(t)$ so that these boundary conditions 
are satisfied.

\begin{center}
\begin{figure}[t]
\begin{center}
\includegraphics[width=0.8\columnwidth]{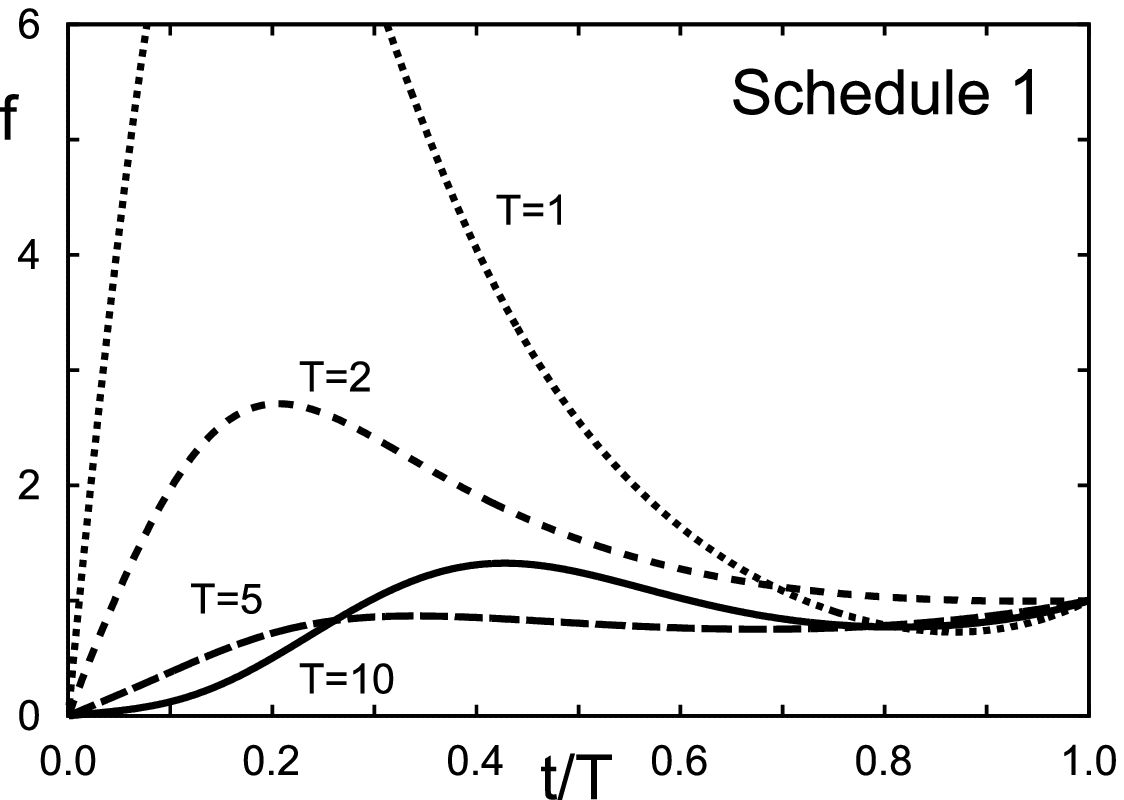}
\includegraphics[width=0.8\columnwidth]{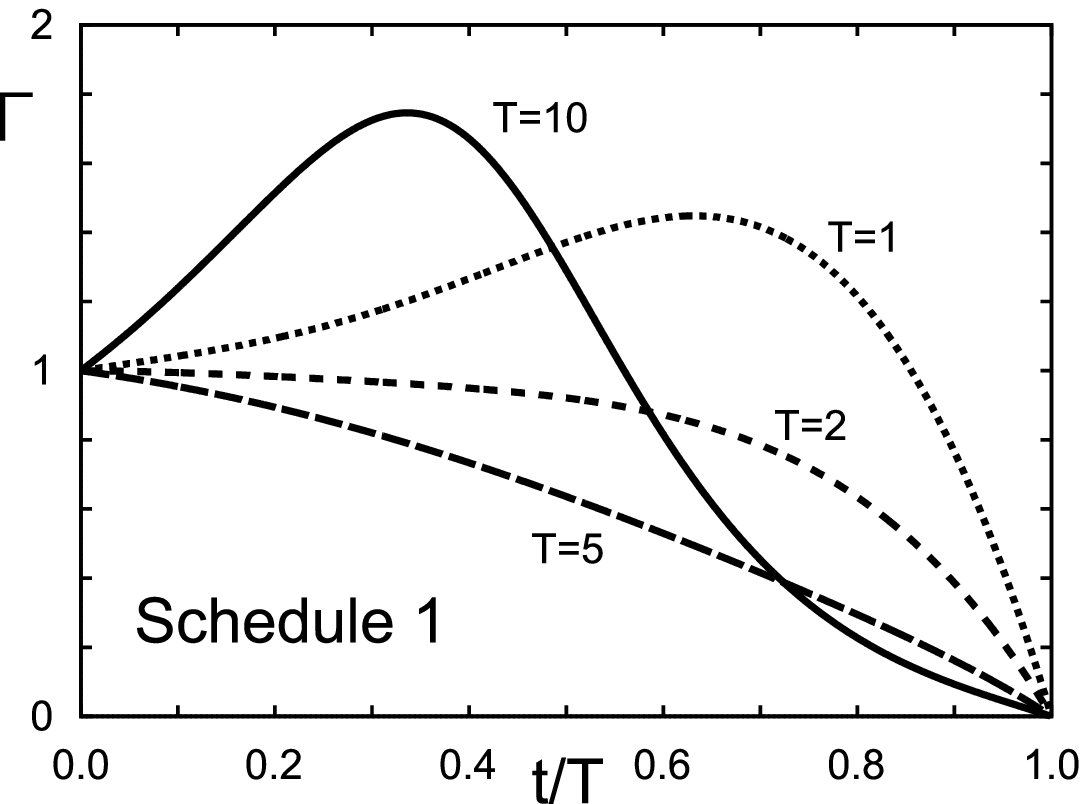}
\end{center}
\caption{Time dependence of the coefficients of the transverse Ising model
(Schedule 1).
}
\label{fig1}
\end{figure}
\end{center}
\begin{center}
\begin{figure}[t]
\begin{center}
\includegraphics[width=0.8\columnwidth]{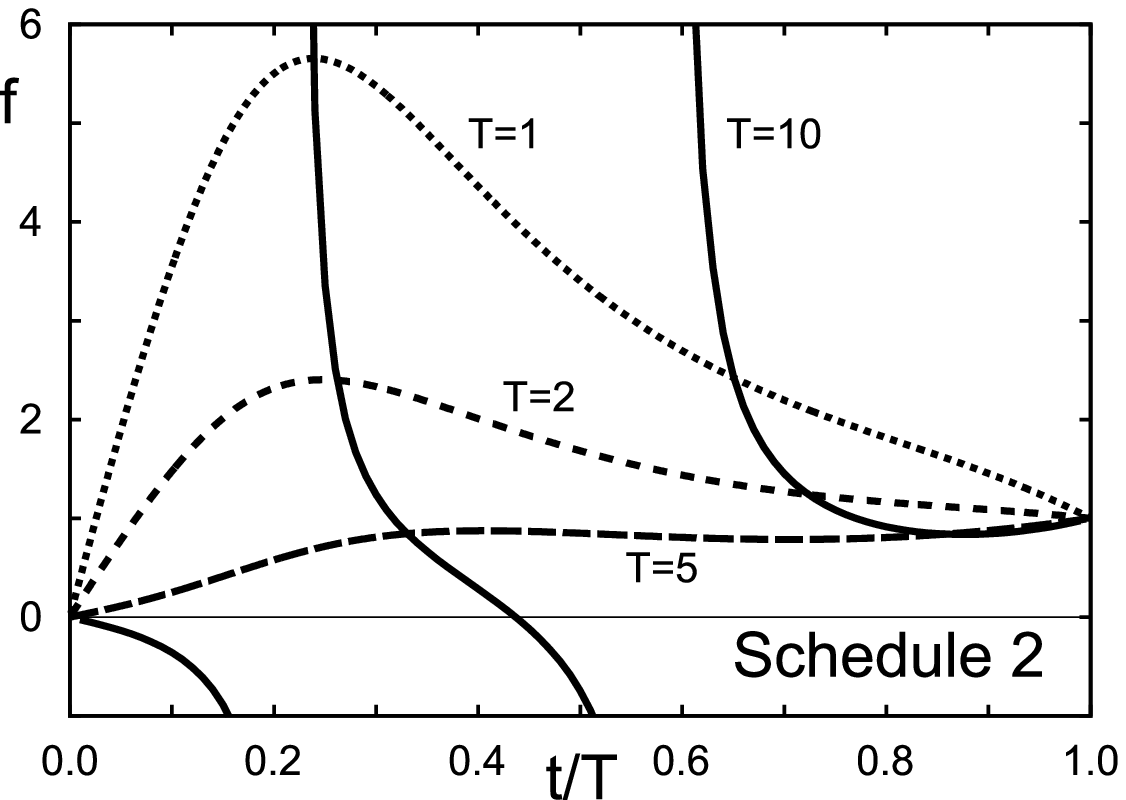}
\includegraphics[width=0.8\columnwidth]{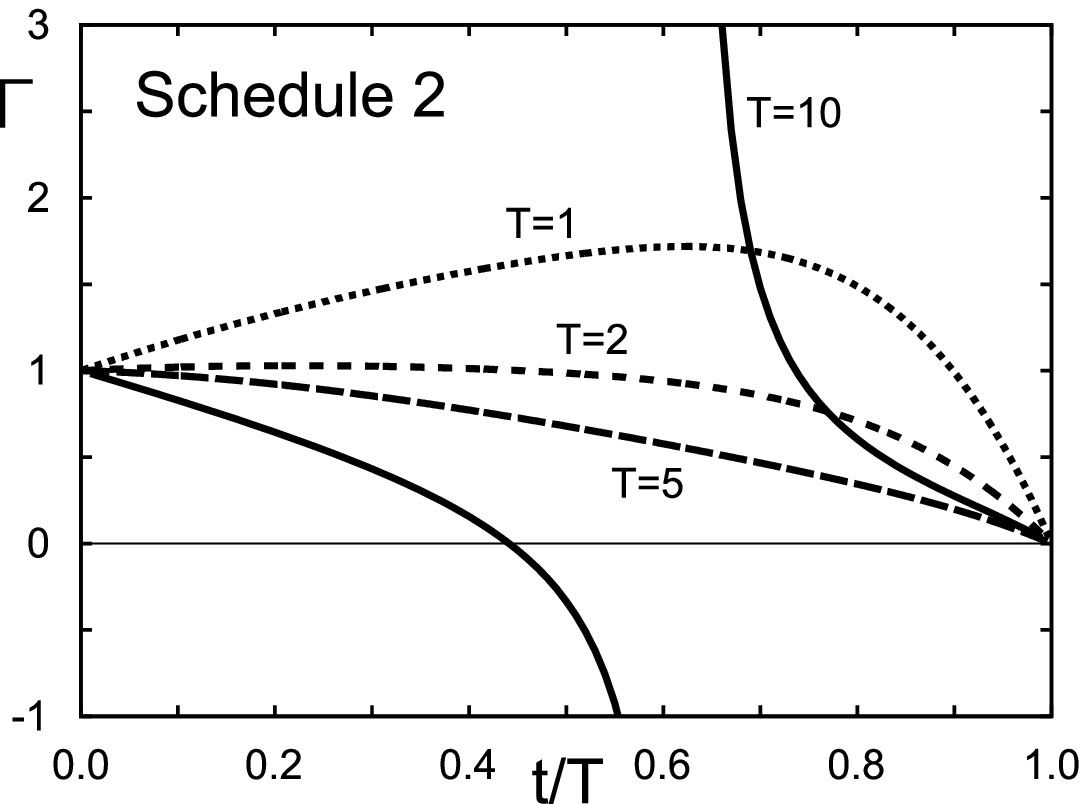}
\end{center}
\caption{Time dependence of the coefficients of the transverse Ising model
(Schedule 2).
}
\label{fig2}
\end{figure}
\end{center}

The calculation is performed in Appendix~\ref{app:mf} 
and we derive two schedules, which we call Schedules 1 and 2. 
The corresponding $\Gamma(t)$ and $f(t)$ are plotted in Fig.~\ref{fig1} 
for Schedule 1 and in Fig.~\ref{fig2} for Schedule 2, respectively. 
Here, we set $\Gamma_0=1.0$ and $h=0.1$.
Since the system is reduced to the two-level systems 
by the mean-field ansatz, 
the schedule is very similar to that in the previous section.
The present schedules do not work well for large $T$ due to 
the divergence of the coefficients.
We also see that $\Gamma$ and $f$ become very large for small $T$.
Thus, we need an intermediate $T$ for practical calculations.

We note that the longitudinal magnetic field $h$ must be finite 
at the initial time $t=0$ since the denominator becomes very small 
at $t=0$, where $\cos\theta(0)=0$.
The finite magnetic field indicates that the state 
avoids the quantum phase transitions throughout the time evolution.
For equilibrium systems with $h=0$, 
the system is in a paramagnetic phase at $t=0$ 
and in a ferromagnetic one at $t=T$.
The energy gap goes to zero and 
the adiabatic approximation fails at the phase boundary.
Furthermore, it is also known that 
the entanglement becomes large and 
the mean-field ansatz is not justified 
around the transition point~\cite{OL, ODV}. 
However, the quantum phase transition is smeared out by 
the presence of the magnetic field $h$.
This property for static systems is also applied to our dynamical systems.
In Sec.~\ref{sec:rot}, we solve this problem in a different way 
by considering a rotating magnetic field.

\subsection{Numerical results}

It is not clear whether the mean-field ansatz is justified
for the the present dynamical system.
We numerically solve the time-dependent Schr\"odinger equation 
without using the mean-field ansatz 
to see the time evolution of the state.

Since the Hamiltonian (\ref{H}) commutes with 
the square of the total spin, 
\be
 \hat{\bm{S}}^2=\left(\sum_{i=1}^N\frac{1}{2}\hat{\bm{\sigma}}_i\right)^2,
\ee
the Hamiltonian is block diagonalized, and each block is specified 
by the eigenvalue $\hat{\bm{S}}^2=S(S+1)$.
In the present choice of the initial condition,
the state is in the sector $S=N/2$ and the time evolution is described by 
the $(N+1)\times(N+1)$ matrix.

\begin{center}
\begin{figure}[t]
\begin{center}
\includegraphics[width=0.8\columnwidth]{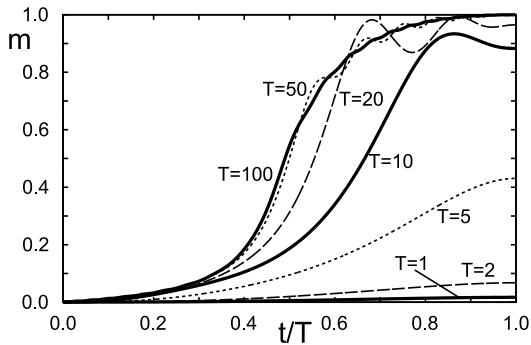}
\end{center}
\caption{Magnetization for linear schedule.
We take $N=4000$, $\Gamma_0=1.0$, and $h=0.1$.}
\label{fig3}
\end{figure}
\end{center}

We numerically solve the problem at $N=4000$.
In the case of linear schedule
\be
 && \Gamma(t) = \Gamma_0\left(1-\frac{t}{T}\right), \\
 && f(t) = \frac{t}{T}, 
\ee
which is the standard schedule for the quantum annealing, 
the time evolution of the magnetization in $z$ direction, 
\be
 m(t)=\frac{1}{N}\sum_{i=1}^N\langle\psi(t)|\hat{\sigma}_i^z|\psi(t)\rangle,
\ee
is plotted in Fig.~\ref{fig3}.
We see that the linear schedule needs a large $T$ to find 
an appropriate evolution.

\begin{center}
\begin{figure}[t]
\begin{center}
\includegraphics[width=0.8\columnwidth]{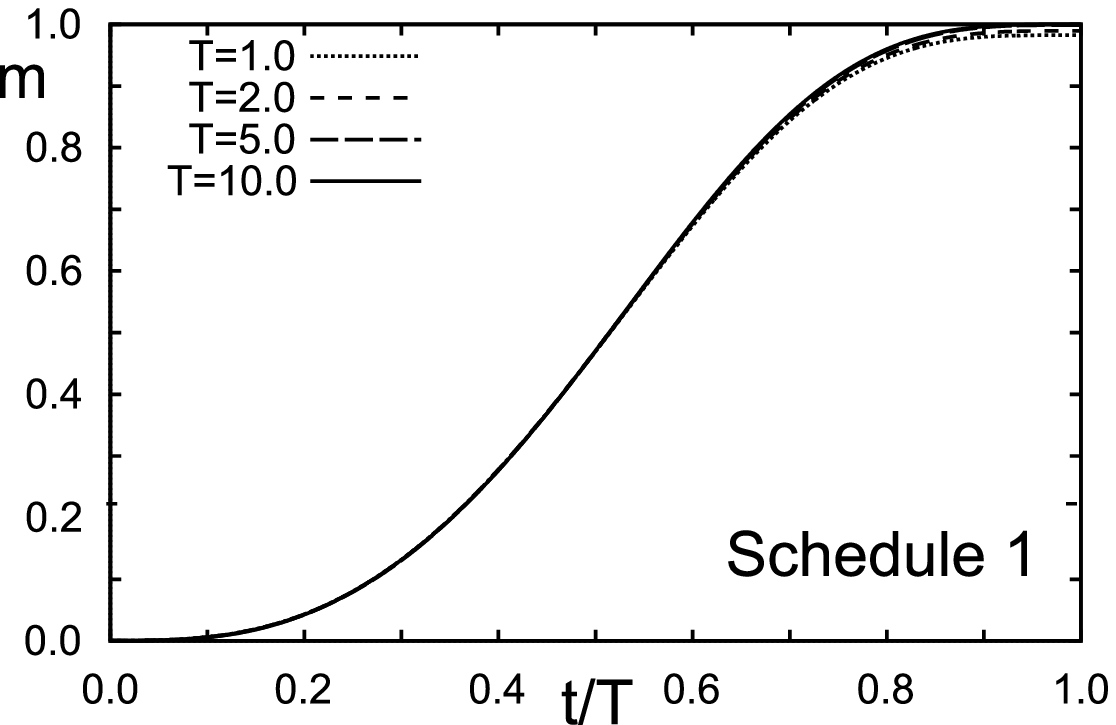}
\includegraphics[width=0.8\columnwidth]{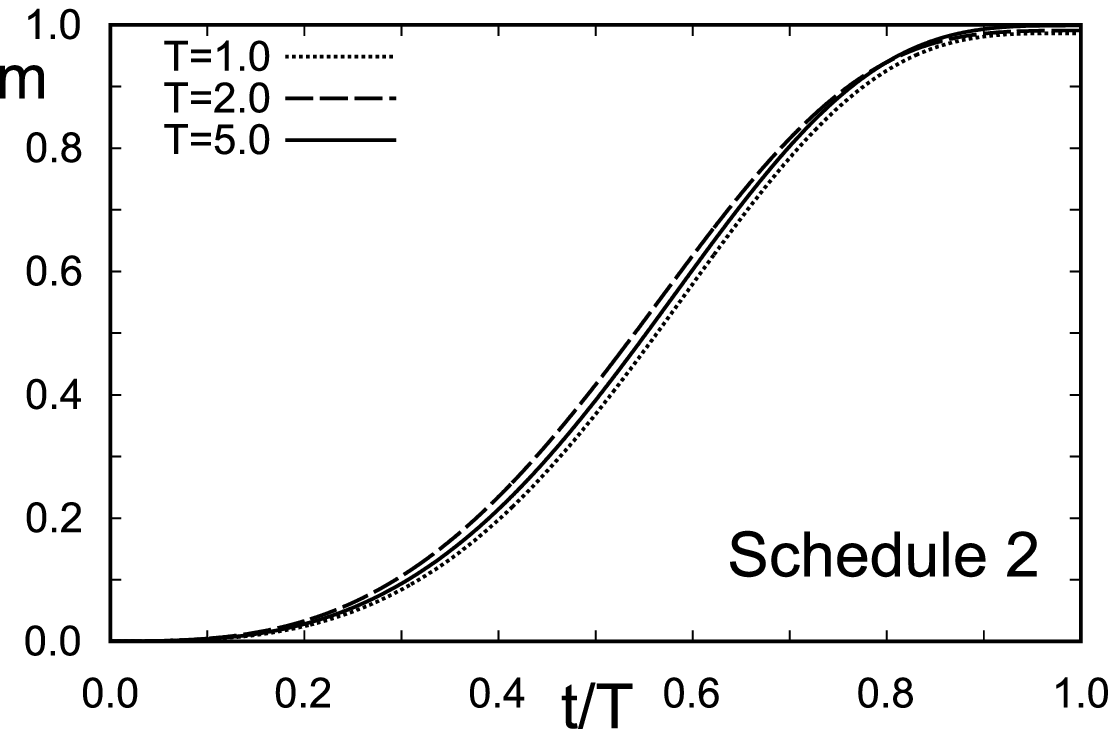}
\end{center}
\caption{Magnetization for Schedules 1 (top) and 2 (bottom).
We take $N=4000$, $J=1.0$, $\Gamma_0=1.0$, and $h=0.1$.}
\label{fig4}
\end{figure}
\end{center}

\begin{center}
\begin{figure}[t]
\begin{center}
\includegraphics[width=0.8\columnwidth]{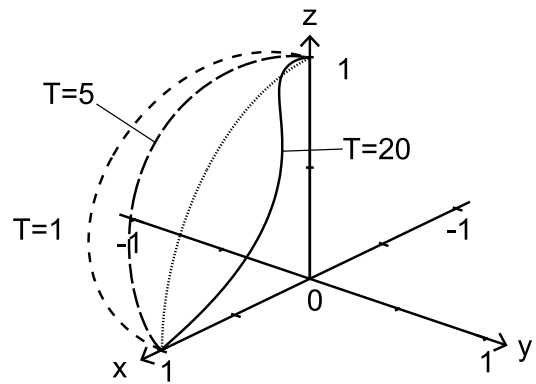}
\end{center}
\caption{Bloch vector for Schedule 1.
}
\label{fig5}
\end{figure}
\end{center}

This behavior is significantly improved by using our schedule
developed in the previous subsection.
As we see in Fig.~\ref{fig4}, 
the results are independent of $T$.
This is the main advantage of our method.
We note that the system follows a different adiabatic passage by using
our schedule, as we see in Fig.~\ref{fig5}.
The Bloch vector follows an adiabatic passage which 
is not in the $zx$ plane and is strongly dependent on $T$.


\subsection{Stability}

We find that the inverse engineering works well
for the mean-field system.
Although the time evolution is independent of the schedule,
the schedule is strongly dependent on the speed of the variation.
In this section, we study the stability of the system
by considering small fluctuations.

The stability of the adiabatic passage was generally discussed 
in Ref.~\cite{Takahashi3}.
For single-spin systems, we can define three directions of perturbations.
The first one is the direction of the Bloch vector: 
\be
 \bm{n}=\bmat{c}
 \sin\theta\cos\varphi \\
 \sin\theta\sin\varphi \\
 \cos\theta
 \emat.
\ee
The Hamiltonian is written by $\bm{n}$ and a different vector: 
\be
 \bm{n}_0=\bmat{c}
 \sin\theta \\ 0 \\ \cos\theta
 \emat.
\ee
It was shown that the adiabatic passage is stable against perturbations
in these two directions.
The remaining third direction is shown to be unstable.
It is given by 
\be
 \frac{\bm{n}\times\bm{n}_0}{|\bm{n}\times\bm{n}_0|} 
 = \frac{1}{\sqrt{\cos^2\theta+\sin^2\theta\cos^2\frac{\varphi}{2}}}
 \bmat{c}
 \cos\theta\cos\frac{\varphi}{2} \\
 \cos\theta\sin\frac{\varphi}{2} \\
 -\sin\theta\cos\frac{\varphi}{2}
  \emat. \no\\
\ee
In the following, we consider $\varphi=0$ for simplicity.
This means that we consider perturbations in the $zx$ plane.

\begin{center}
\begin{figure}[t]
\begin{center}
\includegraphics[width=0.8\columnwidth]{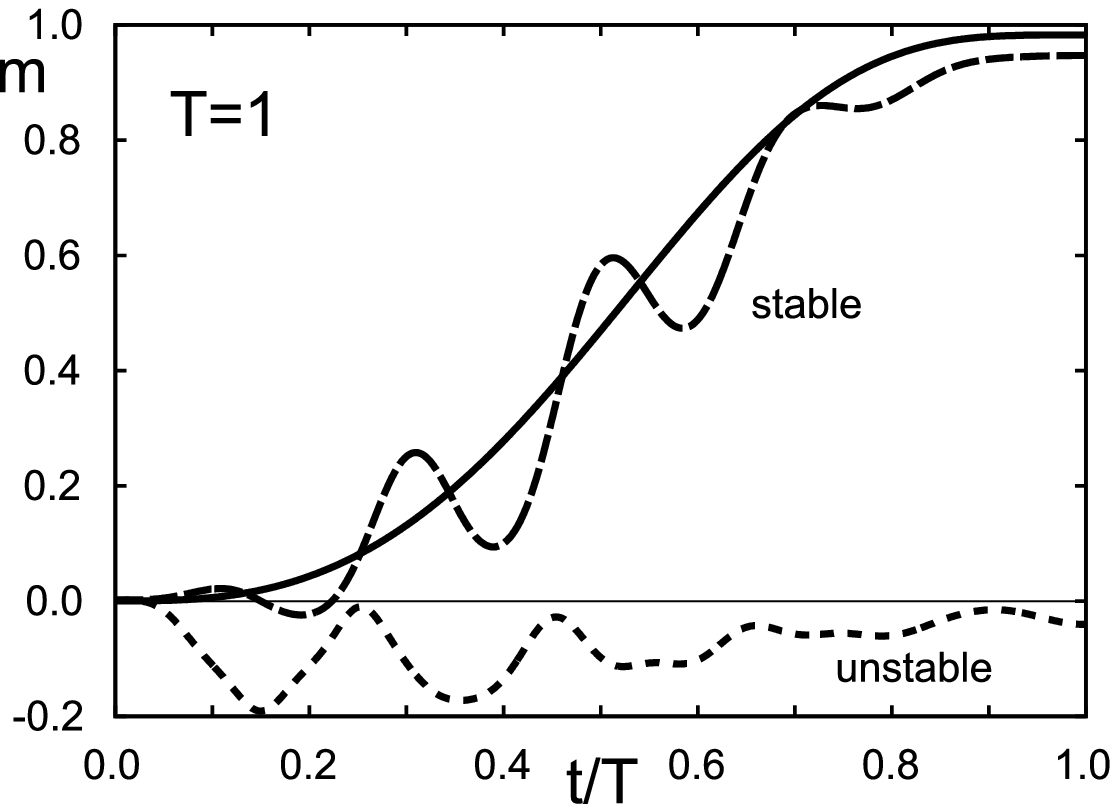}
\includegraphics[width=0.8\columnwidth]{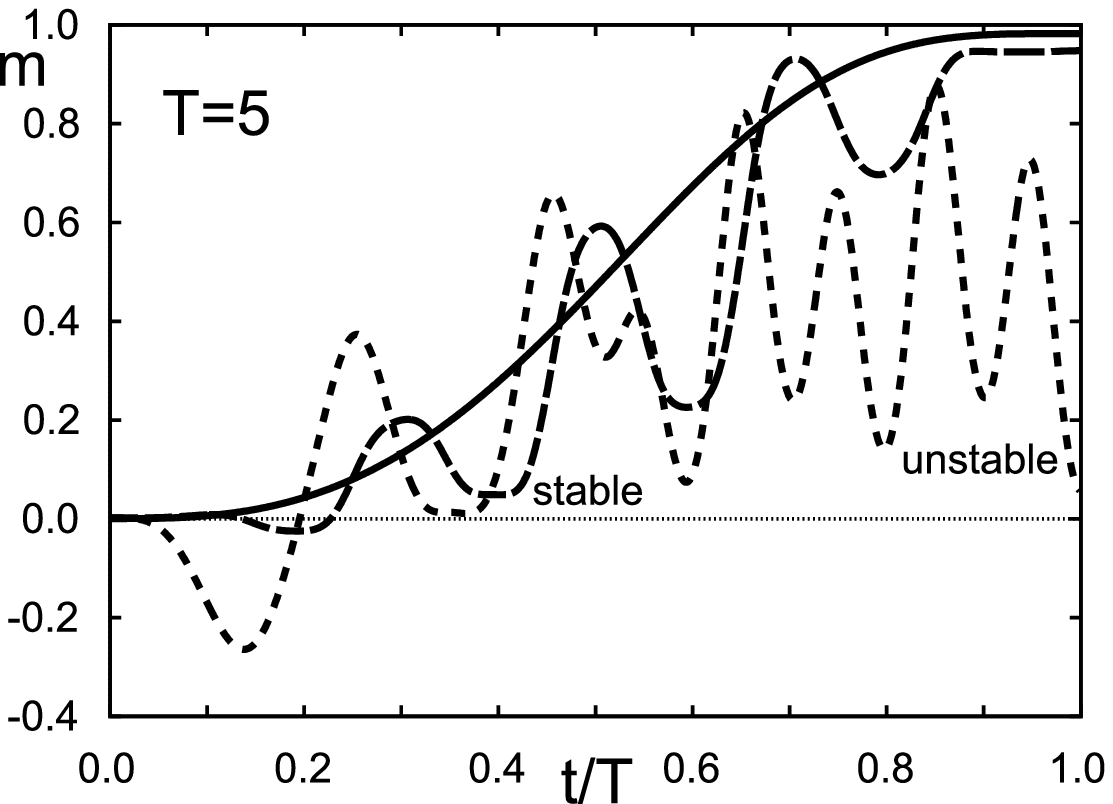}
\includegraphics[width=0.8\columnwidth]{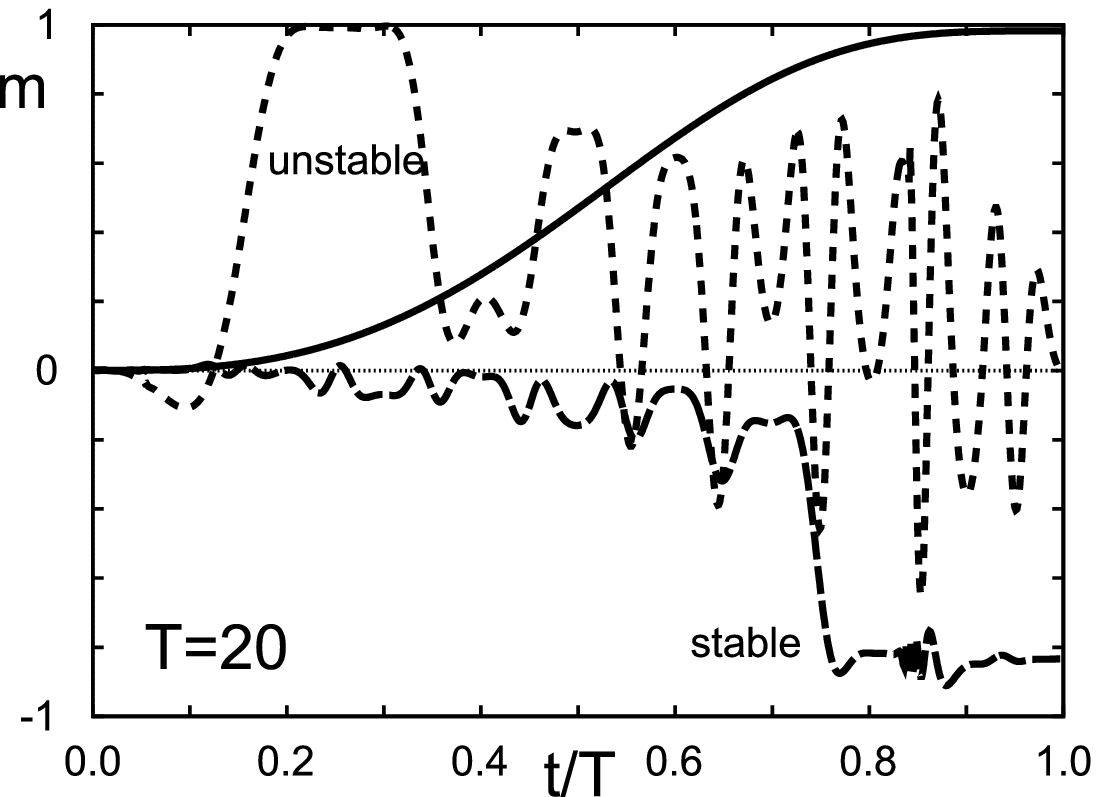}
\end{center}
\caption{Time evolution of the magnetization 
for the Hamiltonian with perturbations.
We set $(h_0,h_p)=(4.0,0.0)$ (stable) and 
$(h_0,h_p)=(0.0,4.0)$ (unstable).}
\label{fig6}
\end{figure}
\end{center}
\begin{center}
\begin{figure}[t]
\begin{center}
\includegraphics[width=0.8\columnwidth]{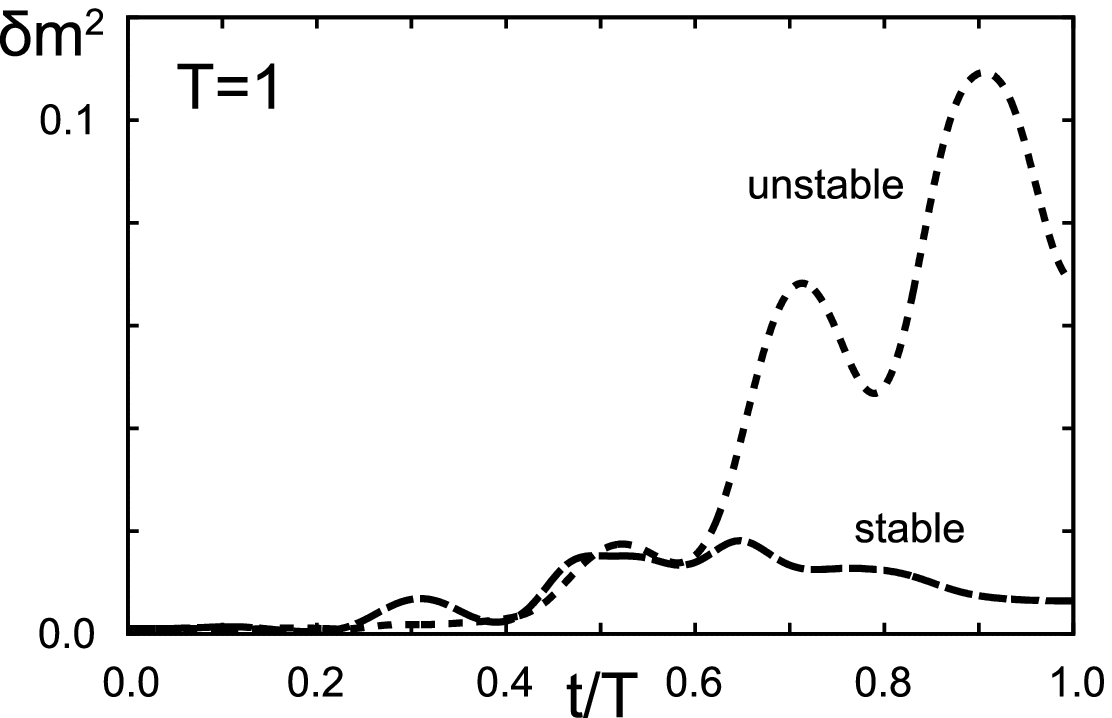}
\includegraphics[width=0.8\columnwidth]{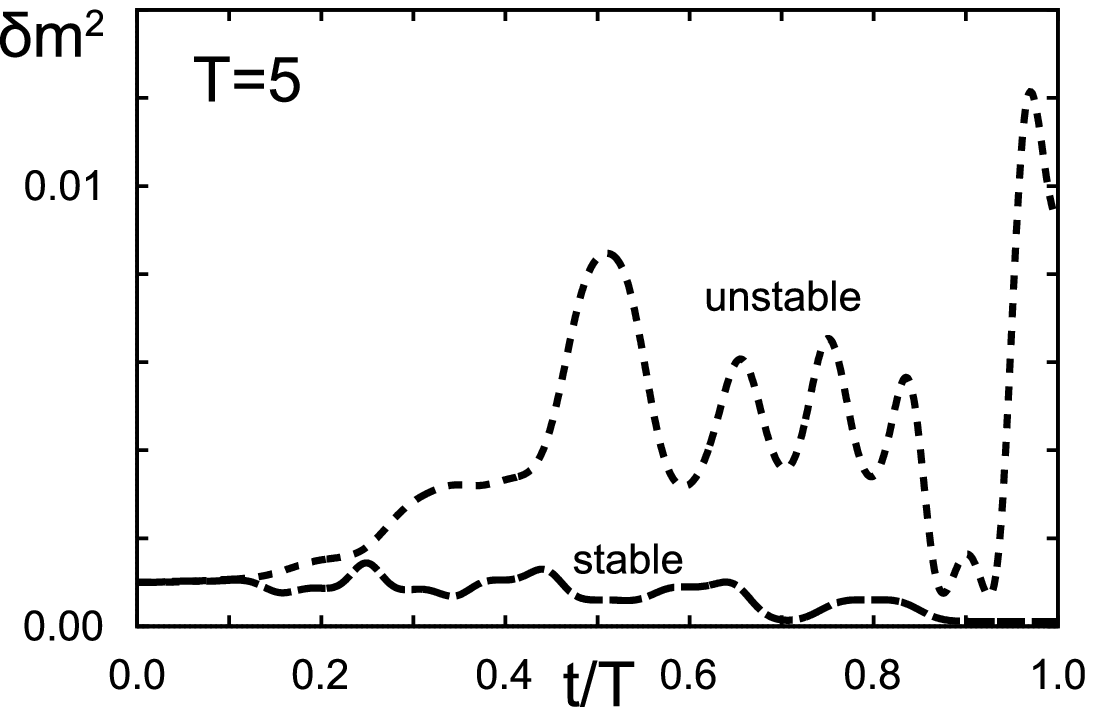}
\includegraphics[width=0.8\columnwidth]{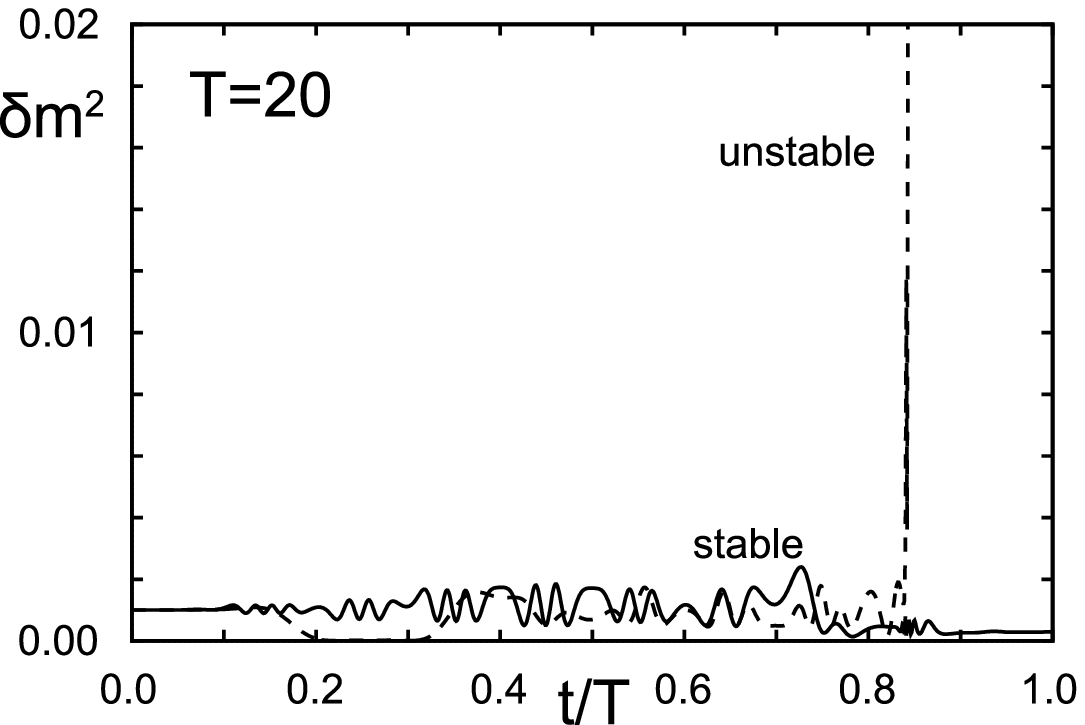}
\end{center}
\caption{Time evolution of the variance of the magnetization 
for the Hamiltonian with perturbations.
We set $(h_0,h_p)=(4.0,0.0)$ (stable) and 
$(h_0,h_p)=(0.0,4.0)$ (unstable).}
\label{fig7}
\end{figure}
\end{center}

Keeping in mind these general considerations, we apply 
the magnetic field: 
\be
 \bm{h}(t)=h_0(t) \bmat{c}
 \sin\theta \\ 0 \\
 \cos\theta
  \emat
 +h_p(t)\bmat{c}
 \cos\theta \\ 0 \\
 -\sin\theta
  \emat.
\ee
We expect that the system is stable against the perturbation $h_0(t)$
and unstable against $h_p(t)$.
We numerically solve the Schr\"odinger equation with the Hamiltonian
$\hat{H}(t)+\bm{h}(t)\cdot\sum_{i=1}^N\bm{\sigma}_i$, 
where $\hat{H}(t)$ is given by Eq.~(\ref{H}).
We use the Schedule 1 and set the magnetic field as 
\be
 && h_0(t)= h_0\sin\frac{\omega t}{T}, \\
 && h_p(t)= h_p\sin\frac{\omega t}{T}.
\ee
We set $\omega=10\pi$ and 
compare the results for $(h_0,h_p)=(4.0,0.0)$ and $(h_0,h_p)=(0.0,4.0)$.
The numerical results are shown in Figs.~\ref{fig6} and \ref{fig7}.
The magnetization is plotted in Fig.~\ref{fig6}
and the variance 
$\delta m^2 = \sum_{i=1}^N \langle\hat{\sigma}_i^z\rangle^2/N$
in Fig.~\ref{fig7}.

We find that the behaviors which are consistent with the general 
argument.
That is, the system is relatively stable against the perturbation $h_0$
and unstable against $h_p$.
Comparing the results with different $T$, 
we also find that an intermediate value of $T$ is most stable 
against perturbations.
We check that similar results are obtained by using several choices of 
$h_0$ and $h_p$.
Thus, although the inverse engineering works for
arbitrary speed of system variation,
we can find the most optimal $T$ by using the stability argument.

\subsection{Rotating magnetic field}
\label{sec:rot}

One of the problem in the inverse engineering
for the transverse Ising model is that a finite magnetic field $h$
is required to find a smooth time evolution.
This is easily understood from the precession motion of spins.
At $t=0$, the direction of the spins is in the positive $x$ axis, 
and the state is insensitive to the magnetic field in the $x$ direction.
This problem is easily circumvented by applying the transverse
field in the $y$ direction.
By introducing a magnetic field in the $xy$ directions, 
we have the $z$-basis Hamiltonian matrix with complex off-diagonal elements.
This is an example of nonstoquastic Hamiltonians.
In this section, we consider the inverse engineering for 
a transverse field in the $xy$ directions.

\begin{center}
\begin{figure}[t]
\begin{center}
\includegraphics[width=0.8\columnwidth]{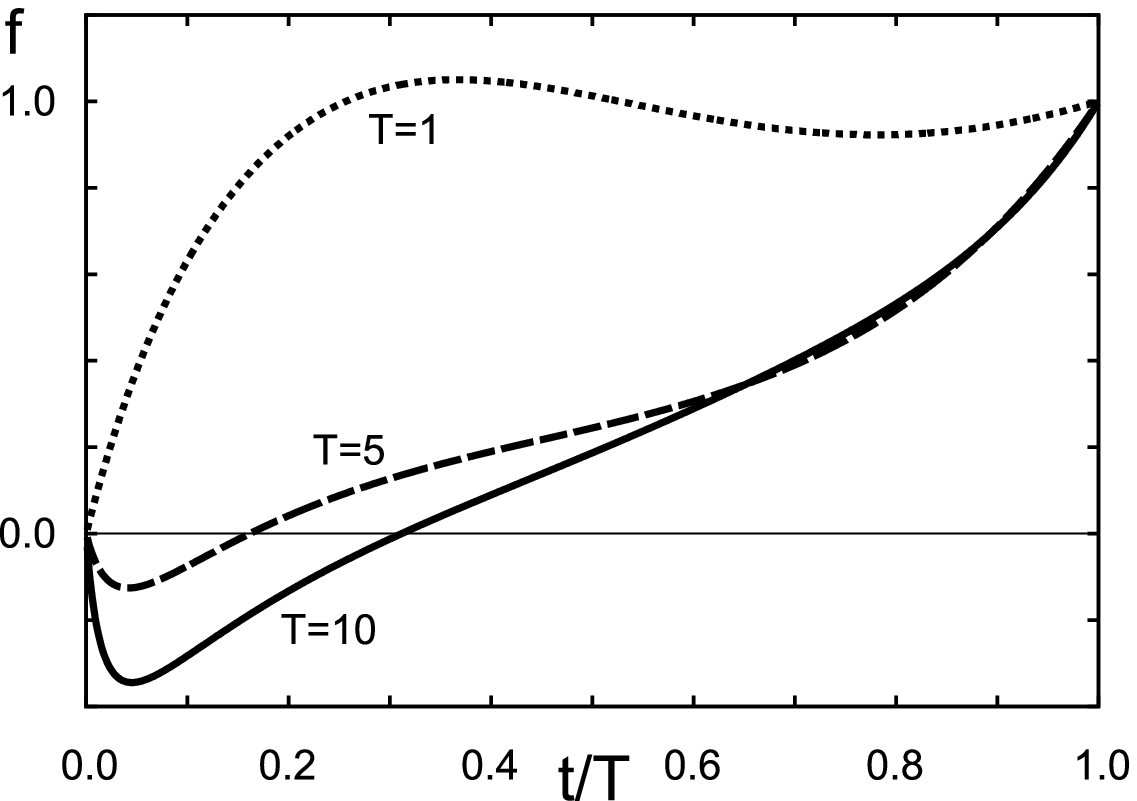}
\includegraphics[width=0.8\columnwidth]{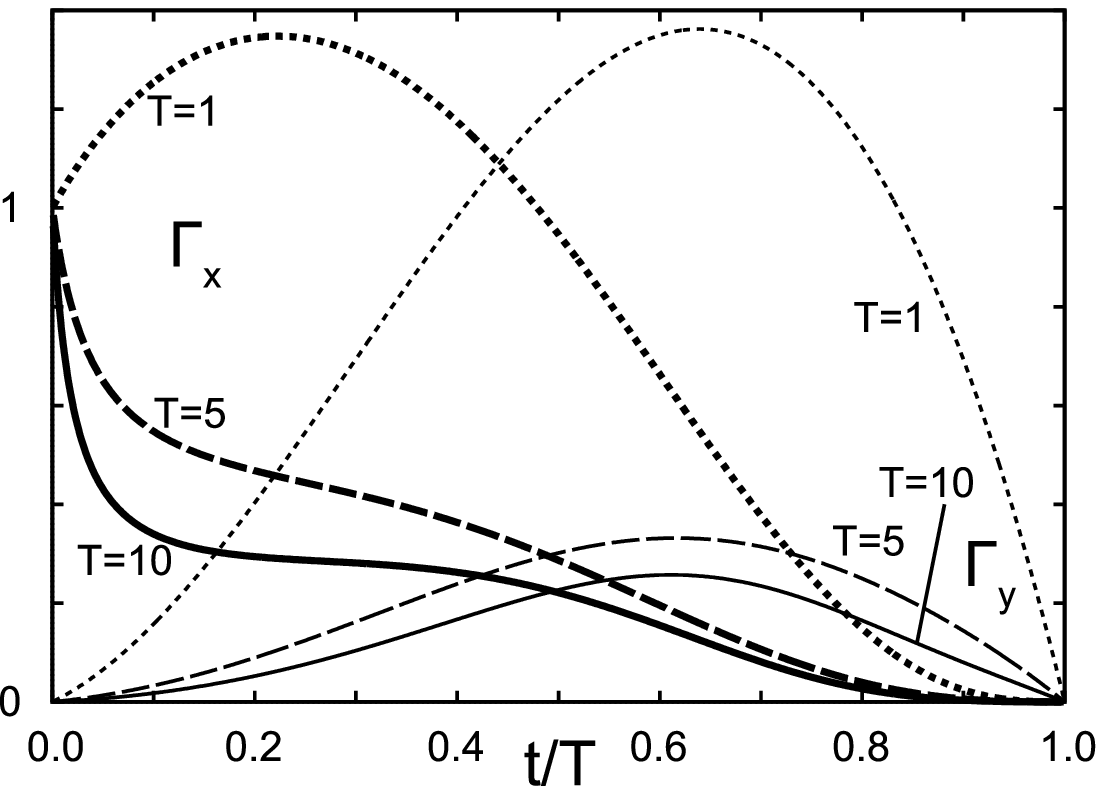}
\end{center}
\caption{Time dependence of the coefficients of the transverse Ising model.
We take $\Gamma_0=1.0$ and $J=1.0$.}
\label{fig8}
\end{figure}
\end{center}

The Hamiltonian is modified as 
\be
 \hat{H}(t) &=&
 -f(t)\frac{J}{2N}\sum_{i,j=1}^N\hat{\sigma}_i^z\hat{\sigma}_j^z\no\\
 && -\Gamma_x(t)\sum_{i=1}^N\hat{\sigma}_i^x
 -\Gamma_y(t)\sum_{i=1}^N\hat{\sigma}_i^y. \label{Hy}
\ee
Following the same calculation as before, we obtain
the equation of motion for the Bloch vector with the mean-field ansatz: 
\be
 \dot{\bm{n}}(t) &\sim& -2f(t)(Jn_z(t)+h)\bm{e}_z\times\bm{n}(t)
 \no\\
 && -2\Gamma_x(t) \bm{e}_x\times\bm{n}(t)
 -2\Gamma_y(t) \bm{e}_y\times\bm{n}(t).
\ee
We parametrize the transverse fields as 
\be
 &&\Gamma_x(t)=\Gamma(t)\cos\gamma(t), \\
 &&\Gamma_y(t)=\Gamma(t)\sin\gamma(t).
\ee
Then, the schedule is determined from the equations
\be
 && \Gamma(t) = \frac{\dot{\theta}(t)}{2\sin(\varphi(t)-\gamma(t))}, \label{gammay}\\
 && f(t) = -\frac{\Gamma(t)}{J\sin\theta(t)\cos\varphi(t)}
 \left[
 \sin(\varphi(t)-\gamma(t))\sin\varphi(t)\right.
 \no\\ && \qquad\quad\left.
 -\cos\gamma(t)\right]
 -\frac{\dot{\varphi}(t)}{2J\cos\varphi(t)}. \label{fy}
\ee
We solve this for the same boundary conditions (\ref{b1}) and (\ref{b2}).
The transverse field in the $y$ direction is zero at $t=0$ and $t=T$.
One of the examples of the angles is given in Appendix~\ref{app:rot}, 
and the corresponding 
schedules are plotted in Fig.~\ref{fig8}.
In this case, the magnetic field in the $z$ direction is not required, 
and we can find a smooth time evolution by a finite magnetic field.

It is important to notice that there is not any problem 
on the gap closing in the present schedule.
As we explained above, the spins change their directions smoothly 
by applying the rotating magnetic field.

\subsection{Generalization}

We have treated the infinite-range model in the above analysis.
The advantage of the model is that 
the mean-field ansatz gives the exact result 
and the ground state of the Hamiltonian with 
all spins aligned in the $z$ direction is easily identified.
In order to apply the present method to general systems
with nontrivial final states, 
we try to apply the analysis to the general Ising model: 
\be
 \hat{H}(t) &=& 
 f(t)\left(
 -\sum_{\langle ij\rangle}J_{ij}\hat{\sigma}_i^z\hat{\sigma}_j^z
 -\sum_{i=1}^N h_i\hat{\sigma}_i^z\right) \no\\
 &&-\Gamma(t)\sum_{i=1}^N\hat{\sigma}_i^x.
\ee
It is a difficult problem to find the ground state of this Hamiltonian. 
Here we study whether 
the mean-field ansatz becomes a general principle to determine the schedule 
and the schedule is determined without knowing the final state.

By assuming the form of the invariant 
\be
 \hat{F}(t)=\sum_{i=1}^N\bm{n}_i(t)\cdot\hat{\bm{\sigma}}_i
\ee
and using the mean-field approximation (\ref{mfa}), 
we obtain
\be
 && \Gamma(t) = \frac{\dot{\theta}_i(t)}{2\sin\varphi_i(t)}, \\
 && f(t) = \frac{1}{2(\sum_j J_{ij}\cos\theta_j(t)+h_i)}
 \no\\
 && \times\left(
 2\Gamma(t)\frac{\cos\theta_i(t)}{\sin\theta_i(t)}\cos\varphi_i(t)
 -\dot{\varphi}_i(t)\right). \label{gammai}
\ee
The left-hand sides of these equations are independent of the site.
We must find appropriate angles $(\theta_i,\varphi_i)$
so that the right-hand sides are independent of $i$.

We assume the initial condition
\be
 (\theta_i(0),\varphi_i(0))=(\pi/2, 0).
\ee
This is the same as before.
As a final condition we assume 
\be
 \theta_i(T)=(1-\xi_i)\frac{\pi}{2} = \left\{\ba{ll}
 0 & \mbox{for}\ \ \xi_i=1 \\
 \pi & \mbox{for}\ \ \xi_i=-1
 \ea\right.,
\ee
where $\xi_i$ denotes the sign of the spin at site $i$.

We assume the ansatz 
\be
 &&\theta_i(t)=\xi_i\theta(t)+(1-\xi_i)\frac{\pi}{2}, \\
 &&\varphi_i(t)=\xi_i\varphi(t).
\ee
Then the equation of motion is written as
\be
 && \Gamma(t) = \frac{\dot{\theta}(t)}{2\sin\varphi(t)}, \\
 && f(t) = \frac{1}{2\left(\sum_j \xi_iJ_{ij}\xi_j\cos\theta(t)+h_i\xi_i\right)}
 \no\\
 && \times
 \left(
 2\Gamma(t)\frac{\cos\theta(t)}{\sin\theta(t)}\cos\varphi(t)
 -\dot{\varphi}(t)\right).
\ee
If we set 
\be
 &&J_{ij}=\frac{1}{N}J\xi_i\xi_j, \label{mattisj}\\
 && h_i =h\xi_i, \label{mattish}
\ee
the equations reduce to the same result as before: 
\be
 && \Gamma(t) = \frac{\dot{\theta}(t)}{2\sin\varphi(t)}, \\
 && f(t) = \frac{1}{2\left(J\cos\theta(t)+h\right)}
 \no\\
 && \times\left(
 2\Gamma(t)\frac{\cos\theta(t)}{\sin\theta(t)}\cos\varphi(t)
 -\dot{\varphi}(t)\right).
\ee
Equations (\ref{mattisj}) and (\ref{mattish}) represents 
the Mattis model~\cite{Mattis}.
This model has no frustration and is reduced to 
the ferromagnetic infinite-range model.
Therefore, the present result is considered to be a natural result.
However, we note that the schedule is independent of $\xi_i$ and 
is determined without knowing the final spin state.

Another possible way to solve the problem is to consider site-dependent
transverse fields. 
We expect that the replacement 
$\Gamma(t)\to \Gamma_i(t)$ can solve Eq.~(\ref{gammai}).
This is an interesting problem but goes beyond the scope of the
present work.
We leave it as a future work.

\section{Conclusions}
\label{sec:conc}

We have studied the inverse engineering of the Hamiltonian 
for quantum annealing.
In the infinite-range model, the mean-field ansatz is used 
to design the schedule, and 
the final state is obtained successfully at arbitrary values of $T$.
This is a nontrivial result even for the infinite-range model 
since we do not know whether the mean-field ansatz gives the exact result
in dynamical systems.
We find that the schedule is strongly dependent on $T$, 
and an intermediate value of $T$ 
is optimal to find realistic values of coefficients and a stable driving.

We note that the schedule is dependent on the final state in principle.
This means that it is impossible to design the Hamiltonian 
without knowing the final state.
However, we showed that it is possible in the Mattis model.
For the general case, the ground state is not unique and 
it is difficult to directly apply the present analysis to general cases.
We expect that there are some approximate ways 
to improve the standard result, 
and this will be our future work.

We also studied the effect of a rotating magnetic field.
In static systems, the direction of the transverse magnetic field 
is not important.
By using a unitary rotation, the Hamiltonian is reduced to 
the standard form with the transverse field in the $x$ direction.
This is also applied to adiabatic dynamics.
In this case, the quantum phase transition occurs at the boundary
between the ferromagnetic phase and paramagnetic phase, 
and the adiabatic approximation fails at this point.
To circumvent a problem, we introduced a rotating magnetic field.
In quantum annealing, the rotation gives dynamical effects 
which cannot be treated statically.
This is easily understood from the precession motion of spins.
Applying the rotating transverse field 
also means the introduction of a nonstoquastic Hamiltonian.
Our result implies that the nonstoquastic Hamiltonian 
is important to find an efficient algorithm.

Finally, we stress the significance of the present method.
The method of shortcuts to adiabaticity 
implies that any system is described by an adiabatic time evolution.
For arbitrary Hamiltonian, the dynamical invariant can 
always be constructed, as we see from an example of the density operator.
Thus, the system follows an adiabatic passage of a reference Hamiltonian.
The problem is that the reference Hamiltonian is not equal to the original
Hamiltonian.
Then the only thing to require is that two trajectories 
coincide at the final time. 
We expect that this picture becomes useful when we understand 
the general properties of quantum time evolutions.

\section*{Acknowledgments}
The author is grateful to Kentaro Imafuku and
Hidetoshi Nishimori for useful discussions.
The author also acknowledges  
financial support from JSPS KAKENHI Grant No. 26400385.

\appendix
\section{Single spin system}
\label{app:single}

\begin{center}
\begin{figure}[t]
\begin{center}
\includegraphics[width=0.8\columnwidth]{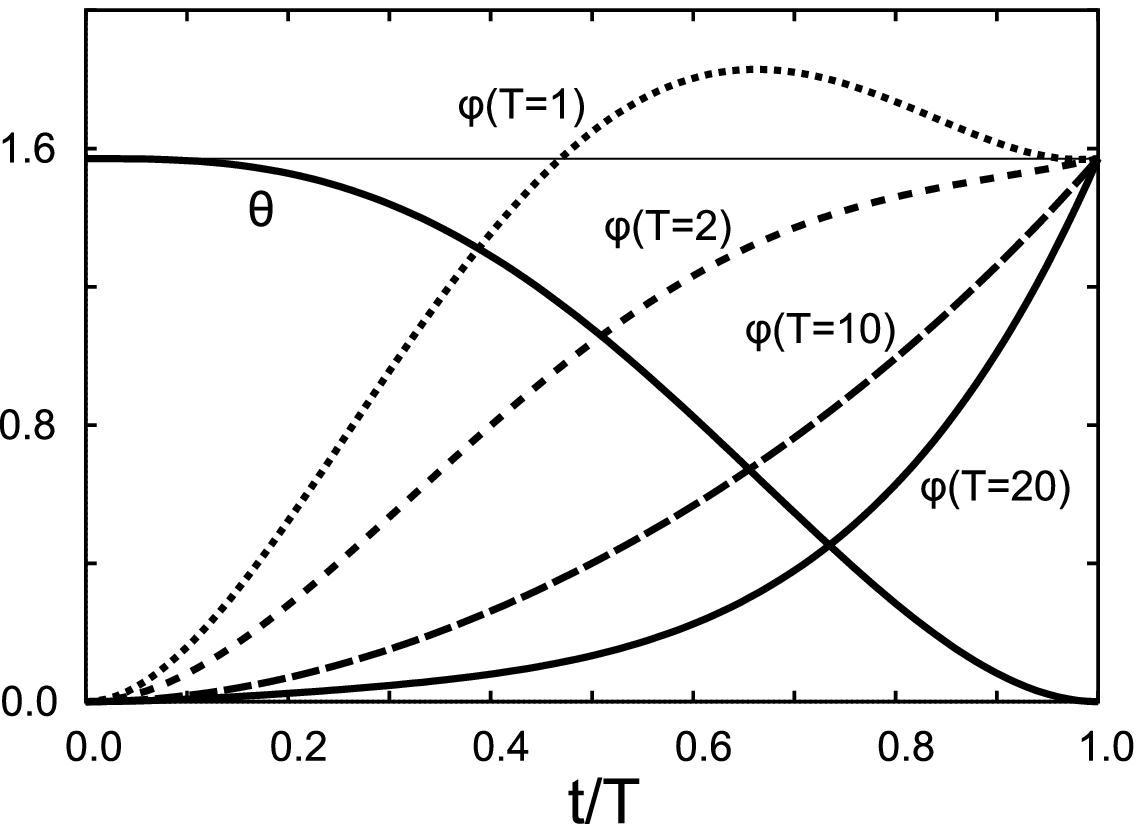}
\includegraphics[width=0.8\columnwidth]{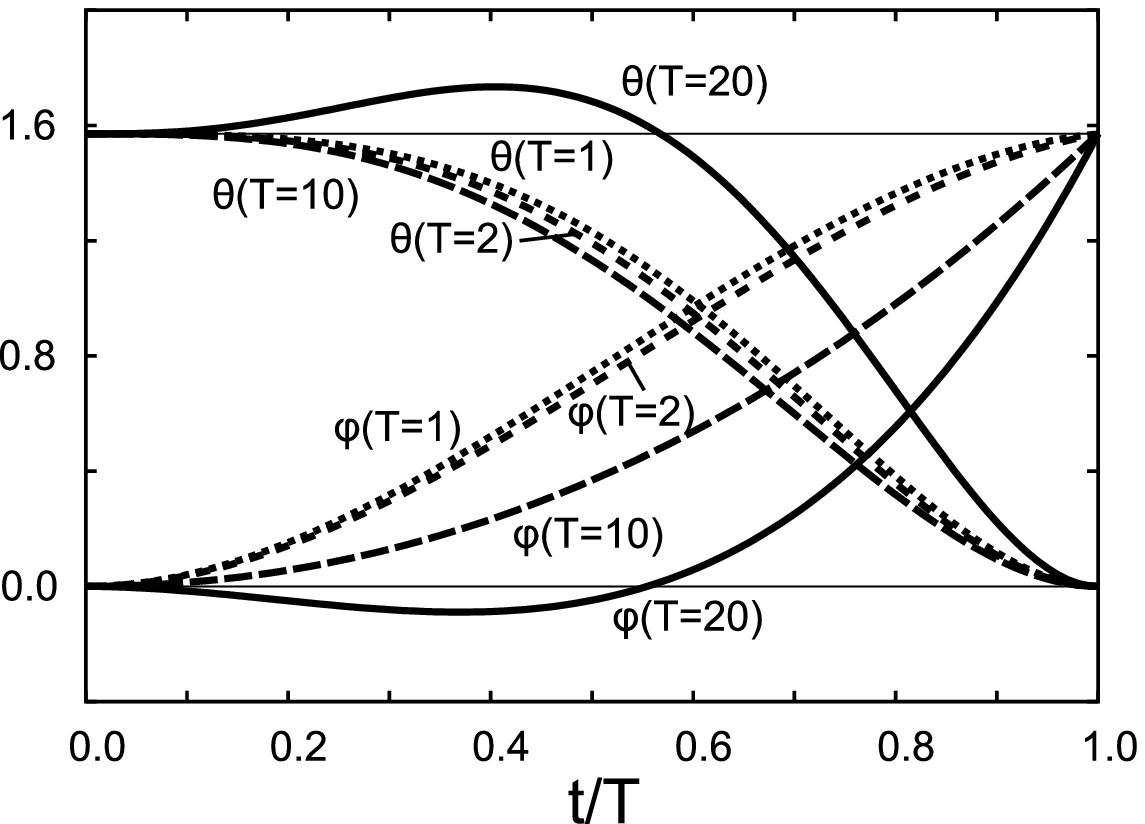}
\end{center}
\caption{Time dependence of angles 
for Schedules 1 (top) and 2 (bottom)
in single-spin system.
We take $\Gamma_0=1.0$ and $h_1=1.0$.}
\label{fig9}
\end{figure}
\end{center}
\begin{center}
\begin{figure}[t]
\begin{center}
\includegraphics[width=0.8\columnwidth]{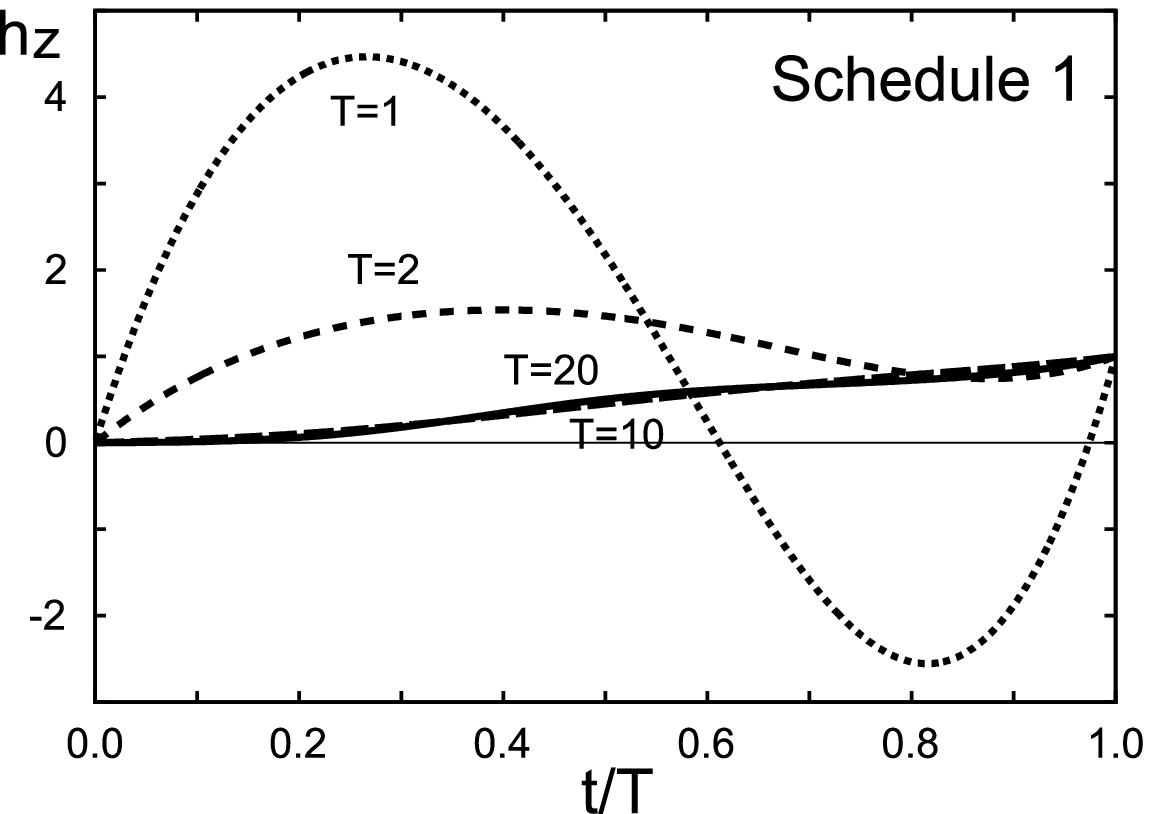}
\includegraphics[width=0.8\columnwidth]{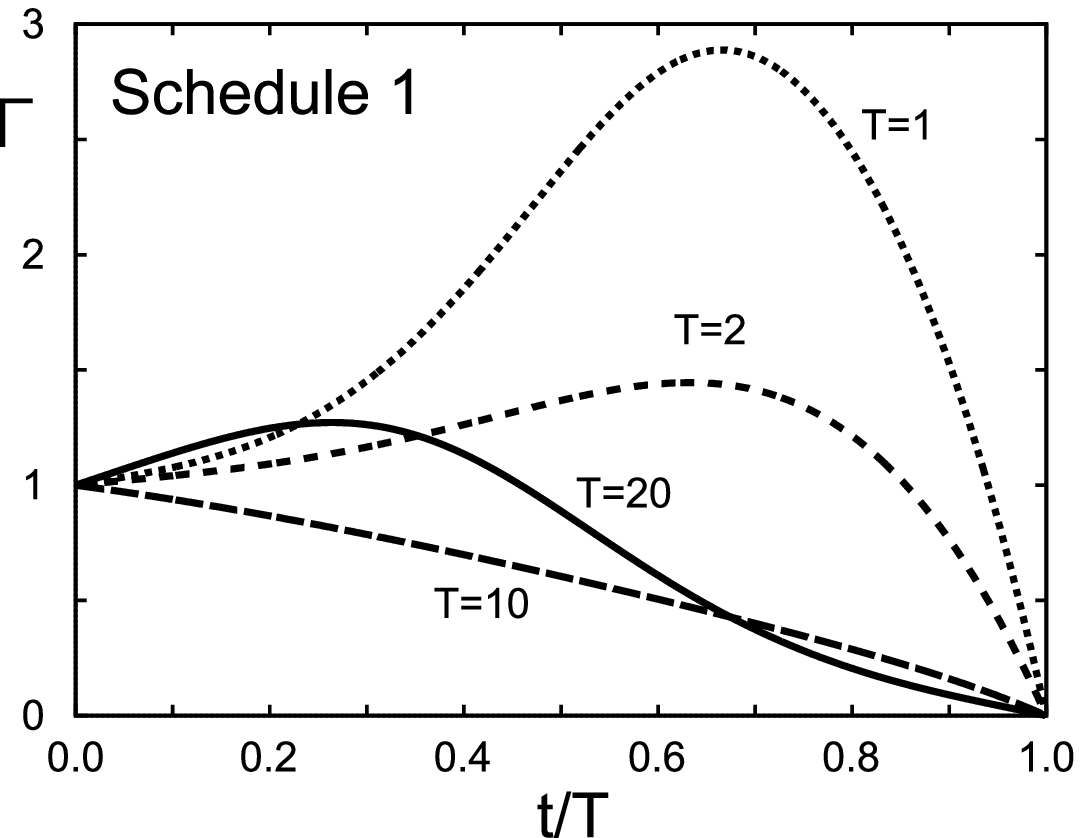}
\end{center}
\caption{Magnetic field for Schedule 1 in the single-spin system.
}
\label{fig10}
\end{figure}
\end{center}
\begin{center}
\begin{figure}[t]
\begin{center}
\includegraphics[width=0.8\columnwidth]{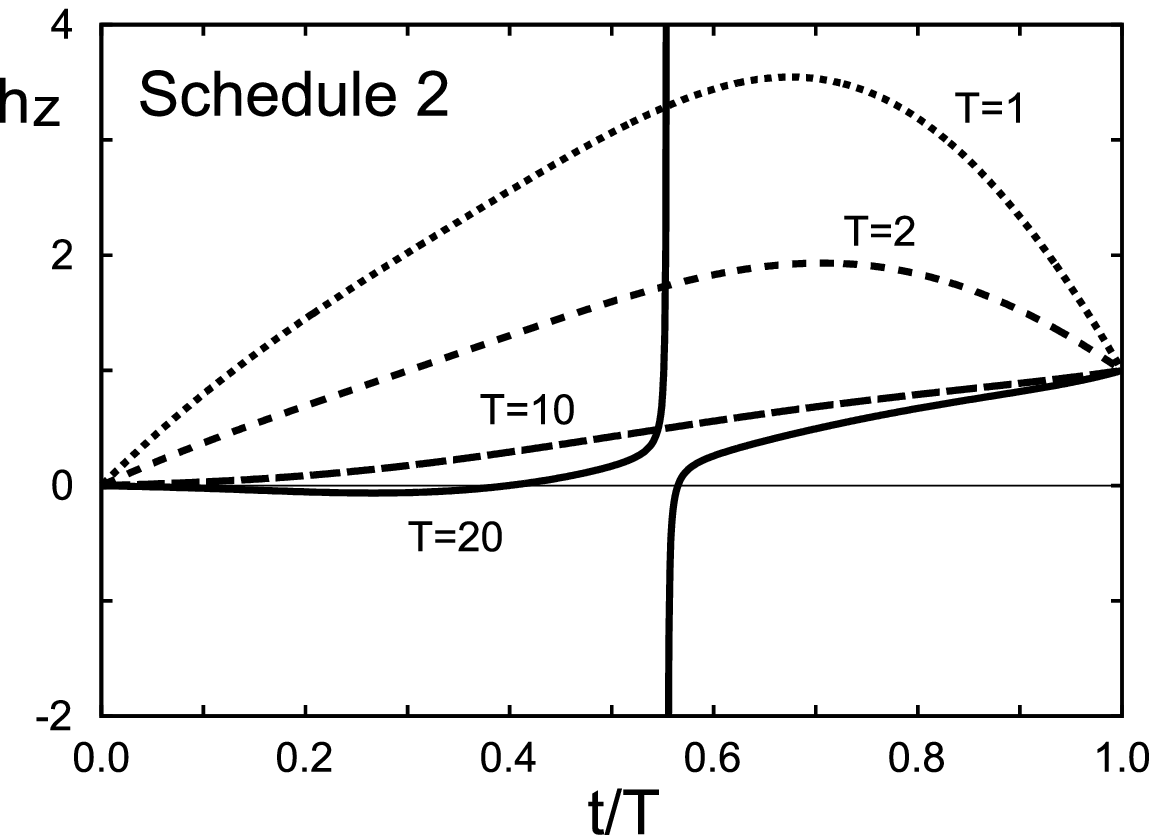}
\includegraphics[width=0.8\columnwidth]{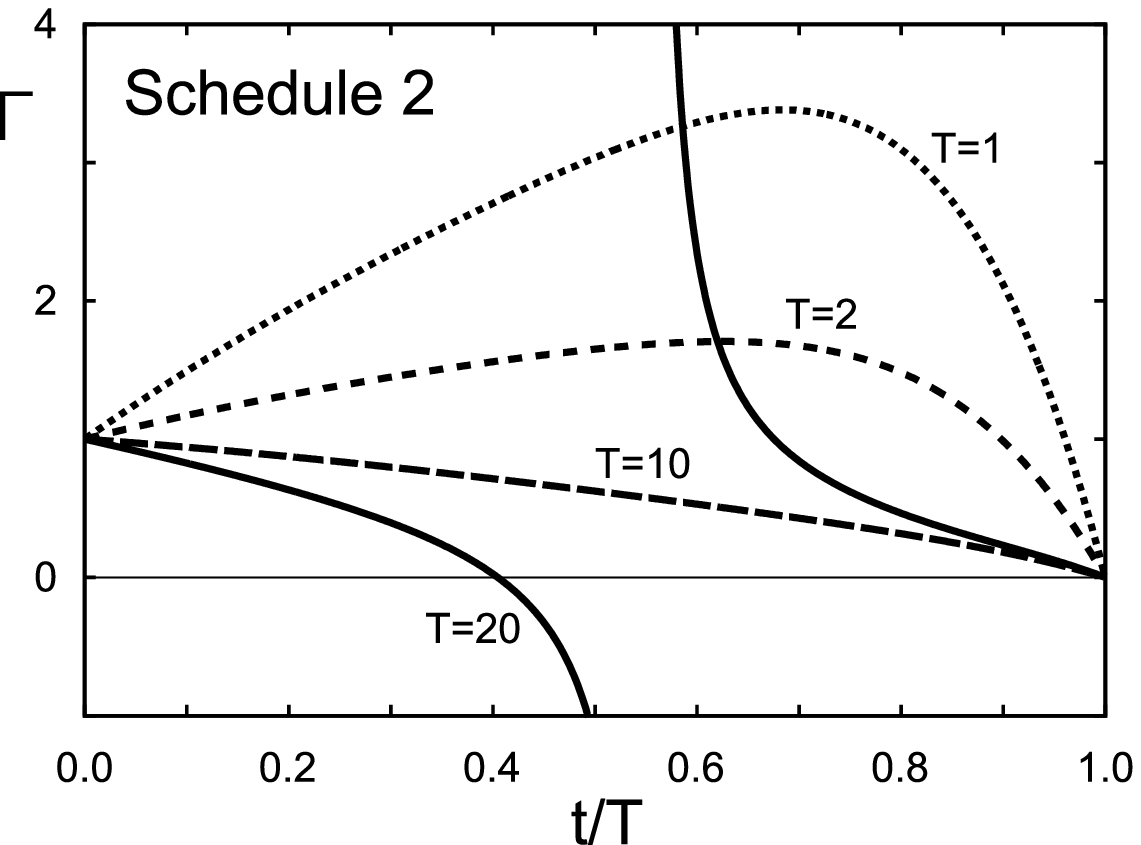}
\end{center}
\caption{Magnetic field for Schedule 2 in the single-spin system.
}
\label{fig11}
\end{figure}
\end{center}

We study the inverse engineering for the single-spin system
in Sec.~\ref{sec:single}.
The magnetic field satisfies the boundary conditions~(\ref{b1}) and (\ref{b2})
and is obtained from Eqs.~(\ref{eq1}) and (\ref{eq2}).
It is represented by angle functions $\theta(t)$ and $\varphi(t)$, 
and they are constrained by the boundary conditions 
(\ref{a-b1})--(\ref{a-b4}).

At $t\to 0$, the angle functions satisfy Eqs.~(\ref{a-b1}) and (\ref{a-b3}).
We expand them as 
\be
 && \theta(t)=\frac{\pi}{2}-\alpha_2\left(\frac{t}{T}\right)^2
 -\alpha_3\left(\frac{t}{T}\right)^3+\cdots, \\
 && \varphi(t)=\beta_2\left(\frac{t}{T}\right)^2
 +\beta_3\left(\frac{t}{T}\right)^3+\cdots.
\ee
Inserting these expressions into Eq.~(\ref{eq1}), we obtain 
\be
 \Gamma \sim \frac{1}{T}\frac{2\alpha_2\frac{t}{T}
 +3\alpha_3\left(\frac{t}{T}\right)^2+\cdots}{
 \beta_2\left(\frac{t}{T}\right)^2
 +\beta_3\left(\frac{t}{T}\right)^3+\cdots}.
\ee
We choose parameters so that this expression goes to $\Gamma_0$ at the limit
$t\to 0$.
We have 
\be
 && \alpha_2=0, \\
 && \beta_2=\frac{3\alpha_3}{\Gamma_0 T}.
\ee
Using Eq.~(\ref{eq2}), we also obtain 
\be
 h_z(t) \sim \Gamma_0 \alpha_3\left(\frac{t}{T}\right)^3
 +\frac{2\beta_2}{T}\frac{t}{T}
 \sim \frac{6\alpha_3}{\Gamma_0 T^2}\frac{t}{T}.
\ee
Since $h_z$ goes to zero at the limit, this is a consistent result.

Next, we study the limit $t\to T$.
We need boundary conditions (\ref{a-b2}) and (\ref{a-b4}).
$\varphi(T)$ is left undetermined.
We put the asymptotic forms as 
\be
 && \theta(t) \sim \gamma_2\left(1-\frac{t}{T}\right)^2
 +\gamma_3\left(1-\frac{t}{T}\right)^3, \\
 && \varphi(t) \sim \varphi(T)-\delta_2\left(1-\frac{t}{T}\right).
\ee
Then, the equation for $\Gamma$ reads 
\be 
 \Gamma \sim \frac{1}{T}\frac{2\gamma_2\left(1-\frac{t}{T}\right)+\cdots}
 {\sin\left[\varphi(T)-\delta_1\left(1-\frac{t}{T}\right)+\cdots\right]}.
\ee
This equation is satisfied at the limit if $\varphi(T)\ne 0$.
For $\varphi(T)= 0$, we need the additional condition $\gamma_2=0$.
The equation for $h_z$ is written as 
\be
 h_z \sim -\frac{\dot{\theta}}{\theta}\frac{\cos\varphi}{\sin\varphi}
 +\dot{\varphi}.
\ee
$\dot{\theta}/\theta$ goes to infinity at the limit, and we need 
$\varphi(T)=\pi/2$ to prevent the divergence.
We thus obtain 
\be
 h_z \sim \frac{3}{T}\delta_1.
\ee

In conclusion, we obtain the asymptotic forms 
\be
 && \theta(t)\sim \left\{\ba{cc}
 \frac{\pi}{2}-\alpha_3\left(\frac{t}{T}\right)^3 & t\sim 0 \\ 
 O\left((1-t/T)^2\right)  & t\sim T \ea\right., \label{cond1}\\
 && \varphi(t)\sim \left\{\ba{cc}
 \frac{3\alpha_3}{\Gamma_0T}\left(\frac{t}{T}\right)^2 & t\sim 0 \\ 
 \frac{\pi}{2}-\frac{h_1T}{3}\left(1-\frac{t}{T}\right)  & t\sim T \ea\right.,
 \label{cond2}
\ee
where $\alpha_3$ denotes a constant.
We set the angles under these boundary conditions.
They are not determined uniquely and we study several simple examples.
\begin{itemize}
\item
Schedule 1: 
\be
 &&\theta(t)
 =\frac{\pi}{2}-2\pi\left(\frac{t}{T}\right)^3+\frac{3\pi}{2}\left(\frac{t}{T}\right)^4, \\
 &&\varphi(t)= \left(\frac{t}{T}\right)^2 \left[
 2\pi\frac{t}{T}\left(1-\frac{3}{4}\frac{t}{T}\right)
 +\frac{6\pi}{\Gamma_0T}\left(1-\frac{t}{T}\right)^2
 \right. \no\\ && \left.
 -\frac{h_1T}{3}\frac{t}{T}\left(1-\frac{t}{T}\right)
 \right].
\ee

\item
Schedule 2: 
\be
 &&\theta(t)=\frac{\pi}{2}
 -\alpha_3\left(\frac{t}{T}\right)^3
 -\alpha_4\left(\frac{t}{T}\right)^4
 -\alpha_5\left(\frac{t}{T}\right)^5,
 \\
 &&\varphi(t)= \left(\frac{t}{T}\right)^2\left[
 \frac{3\alpha_3}{\Gamma_0T}
 +\beta_3\frac{t}{T}
 \right],
\ee
where 
\be
 && \alpha_3 = \frac{\pi}{2}\Gamma_0T-\frac{1}{9}\Gamma_0T h_1T, \\
 && \alpha_4 = \frac{5}{2}\pi-\pi\Gamma_0T+\frac{2}{9}\Gamma_0T h_1T, \\
 && \alpha_5 = -2\pi+\frac{\pi}{2}\Gamma_0T-\frac{1}{9}\Gamma_0T h_1T, \\
 && \beta_3 = -\pi+\frac{h_1T}{3}.
\ee
\end{itemize}
These angles are plotted in Fig.~\ref{fig9}.
The corresponding magnetic fields calculated 
from Eqs.~(\ref{eq1}) and (\ref{eq2}) are shown in
Fig.~\ref{fig10} for Schedule 1 and 
Fig.~\ref{fig11} for Schedule 2.

When $T$ is small, the state changes very quickly, and  
we need large magnetic fields to control the system.
In the other limit of large $T$, the state changes adiabatically.
For Schedule 2 with large $T$, $\varphi$ goes to zero at an intermediate time
and the corresponding magnetic field diverges, as we see from 
Eqs.~(\ref{eq1}) and (\ref{eq2}).
Thus, to find a proper time evolution, 
we need to take a value of $T$ which is not too small 
and to choose the angles so that the vector $\bm{n}(t)$ 
does not cross the $zx$ plane.
We note that the latter problem for Schedule 2 is
due to the power-law ansatz and 
can be circumvented by choosing the schedule in a proper way.

\section{Transverse Ising model}
\label{app:mf}

\begin{center}
\begin{figure}[t]
\begin{center}
\includegraphics[width=0.8\columnwidth]{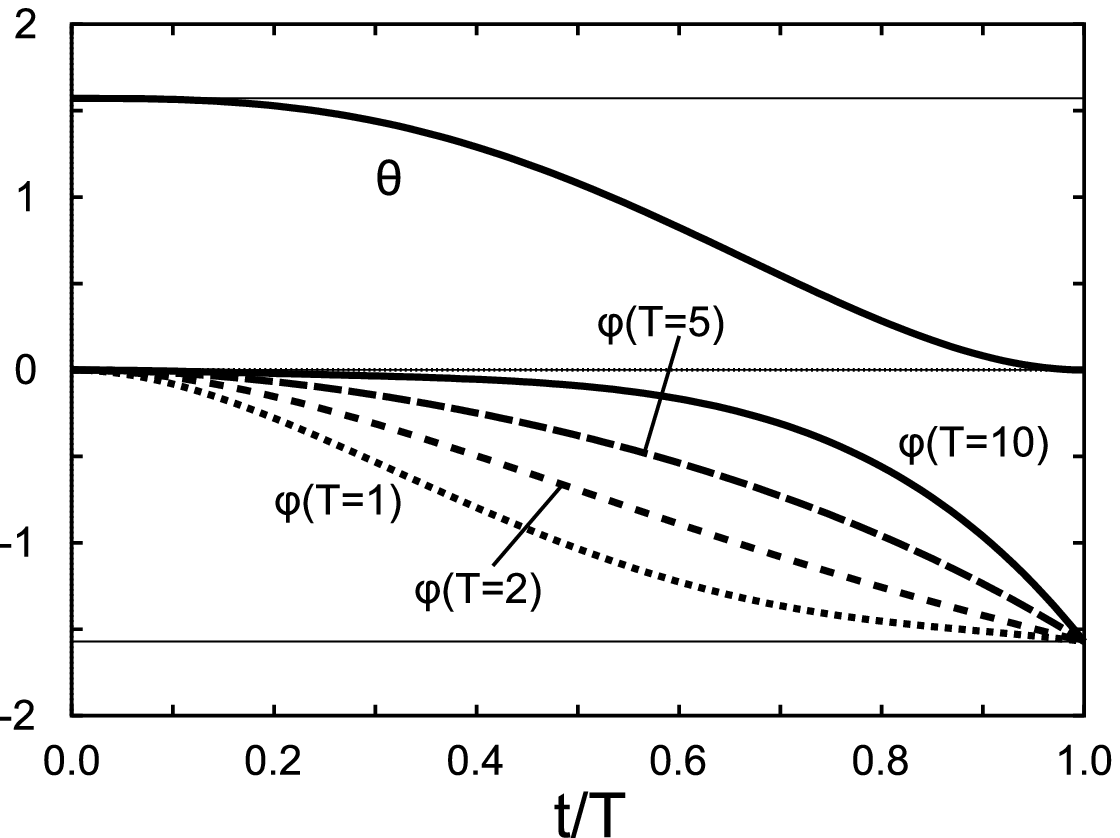}
\includegraphics[width=0.8\columnwidth]{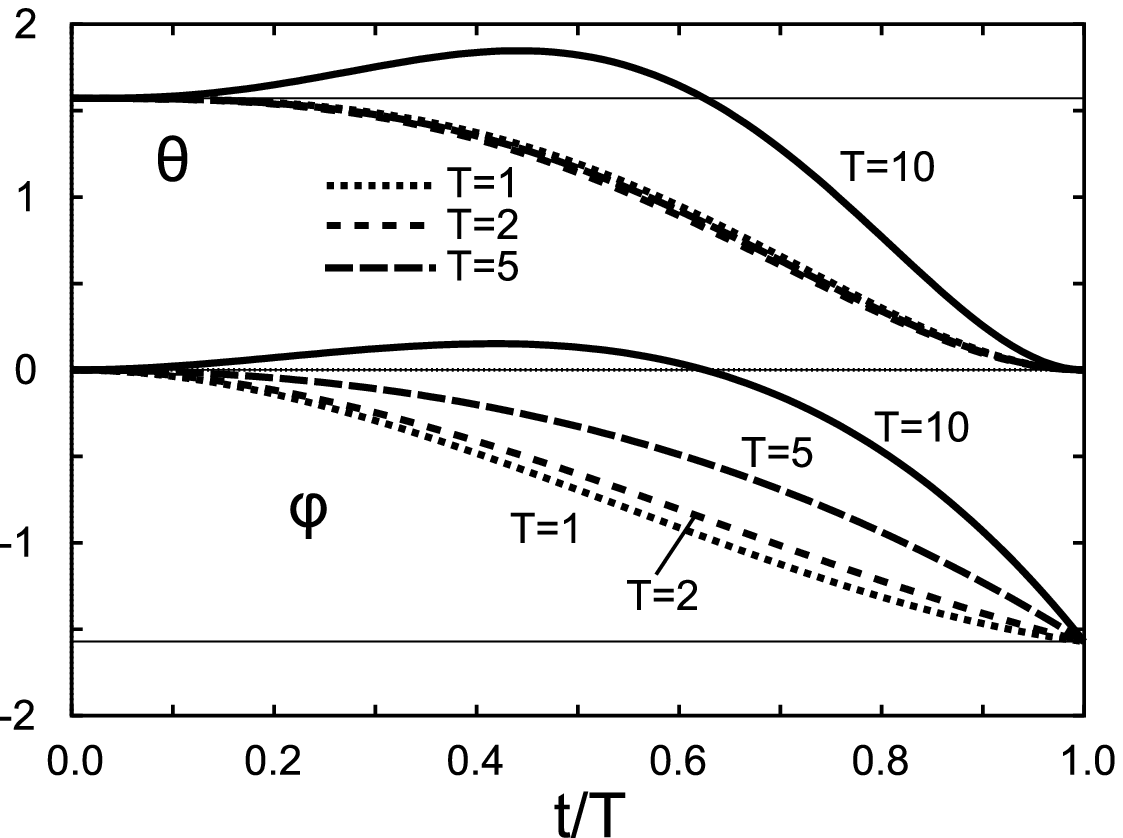}
\end{center}
\caption{Time dependence of angles 
for Schedules 1 (top) and 2 (bottom)
in transverse field Ising model (\ref{H}).
We take $\Gamma_0=1.0$ and $h=0.1$.}
\label{fig12}
\end{figure}
\end{center}

We determine the schedule for the transverse Ising model.
The coefficients of the Hamiltonian are given by 
(\ref{gamma}) and (\ref{f}) 
under the boundary conditions
(\ref{initial}) and (\ref{final}).
We choose angles $\theta(t)$ and $\varphi(t)$ in a proper way.

First, we determine asymptotic forms of the angles.
The calculation goes along the same lines as in the single-spin case.
From Eqs.~(\ref{gamma}) and (\ref{f}), 
the initial condition for the angles is written as 
\be
 &&\theta(t)=\frac{\pi}{2}+\frac{2\Gamma_0}{3}\alpha t^3 +O(t^4), \\
 &&\varphi(t)=\alpha t^2+O(t^3),
\ee
where $\alpha$ is an arbitrary parameter.
In the same way, the final condition at $t=T$ reads
\be
 && \theta(t)=O((T-t)^2), \\
 && \varphi(t)=\frac{\pi}{2}+\frac{2}{3}(J+h)(T-t)+O((T-t)^2).
\ee

We consider the following two schedules.
\begin{itemize}
\item
Schedule 1: 
\be
 &&\theta(t)
 =\frac{\pi}{2}-2\pi\left(\frac{t}{T}\right)^3+\frac{3\pi}{2}\left(\frac{t}{T}\right)^4, 
 \\
 &&\varphi(t)= -\left(\frac{t}{T}\right)^2 \left[
 2\pi\frac{t}{T}\left(1-\frac{3}{4}\frac{t}{T}\right)
 +\frac{3\pi}{\Gamma_0T}\left(1-\frac{t}{T}\right)^2
 \right.\no\\ && \left.
 -\frac{2(J+h)T}{3}\frac{t}{T}\left(1-\frac{t}{T}\right)
 \right].
\ee
\item
Schedule 2: 
\be
 &&\theta(t)=\frac{\pi}{2}
 -\alpha_3\left(\frac{t}{T}\right)^3
 -\alpha_4\left(\frac{t}{T}\right)^4
 -\alpha_5\left(\frac{t}{T}\right)^5,
 \\
 &&\varphi(t)= -\left(\frac{t}{T}\right)^2\left[
 \frac{3\alpha_3}{2\Gamma_0T}
 +\beta_3\frac{t}{T}
 \right], 
\ee
where 
\be
 && \alpha_3 = \pi\Gamma_0T-\frac{4}{9}\Gamma_0T (J+h)T, \\
 && \alpha_4 = \frac{5}{2}\pi-2\pi\Gamma_0T+\frac{8}{9}\Gamma_0T (J+h)T, \\
 && \alpha_5 = -2\pi+\pi\Gamma_0T-\frac{4}{9}\Gamma_0T (J+h)T, \\
 && \beta_4 = -\pi+\frac{2(J+h)T}{3}.
\ee
\end{itemize}
These angles are plotted in Fig.~\ref{fig12}.

\section{Rotating magnetic field}
\label{app:rot}

\begin{center}
\begin{figure}[t]
\begin{center}
\includegraphics[width=0.8\columnwidth]{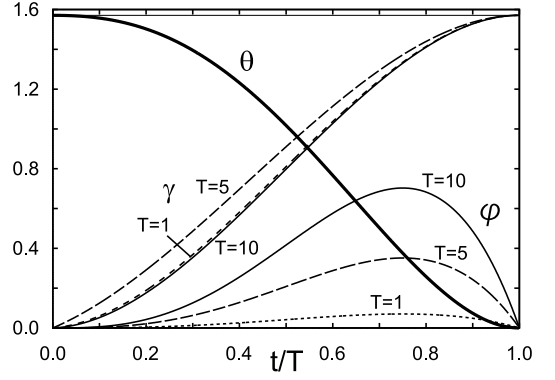}
\end{center}
\caption{Time dependence of angles.
We take $\Gamma_0=1.0$ and $J=1.0$.}
\label{fig13}
\end{figure}
\end{center}

We determine the schedule for the transverse Ising model where 
the magnetic field is in the $xy$ plane.
The coefficients of the Hamiltonian are given by 
(\ref{gammay}) and (\ref{fy}) 
under the boundary conditions
\be
 && (f(0),\Gamma_x(0),\Gamma_y(0))= (0,\Gamma_0,0),\\
 && (f(T),\Gamma_x(T),\Gamma_y(T))= (1,0,0).
\ee
A possible form of the angles is given by 
\be
 && \theta(t)=\frac{\pi}{2}-\frac{J}{\Gamma_0}\left(\frac{t}{T}\right)^2
 +\left(\frac{3\pi}{2}-\frac{2J}{\Gamma_0^2T}\right)\left(\frac{t}{T}\right)^3
 \no\\
 && \quad\qquad +\left(-\pi+\frac{J}{\Gamma_0^2T}\right)\left(\frac{t}{T}\right)^4, \\
 && \varphi(t)=\frac{2}{3}JT\left(\frac{t}{T}\right)^3
 \left(1-\frac{t}{T}\right), \\
 && \gamma(t) = \frac{J}{\Gamma_0^2T}\frac{t}{T}
 +\left(\frac{3\pi}{2}-\frac{2J}{\Gamma_0^2T}\right)\left(\frac{t}{T}\right)^2
 \no\\
 && \quad\qquad +\left(-\pi+\frac{J}{\Gamma_o^2T}\right)\left(\frac{t}{T}\right)^3.
\ee
This is plotted in Fig.~\ref{fig13}.


\end{document}